\RequirePackage{fix-cm}
\documentclass[smallextended]{svjour3}
\usepackage{graphicx}
\usepackage{amsmath, amsfonts, amssymb}
\usepackage{bm}
\usepackage{cite}
\allowdisplaybreaks
\begin{document}
\title{Revisiting the Stress Field Inside an Elastic Sphere Subjected to a Concentrated Load}
\titlerunning{Revisiting the Stress Field Inside an Elastic Sphere}
\author{Yosuke Mori \and Kiwamu Yoshii \and Satoshi Takada}
\institute{
    Y. Mori \at
    Department of Industrial Technology and Innovation, 
    Tokyo University of Agriculture and Technology,
    2-24-16 Naka-cho, Koganei, Tokyo 184-8588, Japan 
    \and
    K. Yoshii \at
    Department of Applied Physics,
    Tokyo University of Science, 
    6-3-1, Niijuku, Katsushika-ku Tokyo 125-8585, Japan\\
    \email{qyoshii@rs.tus.ac.jp}
    \and
    S. Takada \at
    Department of Mechanical Systems Engineering, 
    Tokyo University of Agriculture and Technology,
    2-24-16 Naka-cho, Koganei, Tokyo 184-8588, Japan\\
    Tel.: +81-42-388-7702\\
    Corresponding author, \email{takada@go.tuat.ac.jp}
}

\date{Received: date / Accepted: date}
\maketitle

\begin{abstract}
We present a complete analytical solution for the stress field inside a homogeneous, inside a homogeneous, linearly elastic solid sphere subjected to a concentrated normal load applied on its surface. 
Starting from the three-dimensional linearized elastodynamic equations, the displacement and stress fields are derived using scalar and vector potential representations combined with spherical harmonic expansions. 
All expansion coefficients are determined explicitly by enforcing the traction boundary conditions.
The static elastic solution is obtained rigorously as the long-time limit of the dynamical formulation. 
Closed-form expressions for all components of the stress tensor are provided, enabling direct evaluation of the principal stresses and their differences throughout the interior of the sphere. 
The analytical solution is further generalized to arbitrary loading positions by means of rotational transformations, allowing systematic treatment of multiple concentrated loads through superposition. 
\end{abstract}

\section{Introduction}
Analytical solutions in three-dimensional linear elasticity remain essential for understanding stress concentration, singular behavior, and the internal mechanical response of solids subjected to localized loading~\cite{Timoshenko,Eringen,Aki,Fung}. 
Classical point-force solutions, such as those derived by Kelvin and Boussinesq, have played a foundational role in elasticity theory~\cite{Timoshenko}. 
However, these solutions are restricted to infinite or semi-infinite domains. 
For bounded three-dimensional bodies, particularly those with curved boundaries, exact analytical solutions are considerably fewer~\cite{Hiramatsu66,Jingu85_3D,Sato24_3D} despite their fundamental importance.

A solid sphere subjected to a concentrated surface load constitutes a canonical yet nontrivial boundary-value problem in elasticity~\cite{Timoshenko,Eringen,Fung,Sternberg52,Hiramatsu66,Jingu85_3D,Sato24_3D,Schonert04,Wu06,Ma08,Ramesh22,Shins23}. 
Although spherical geometries under axisymmetric or distributed loading have been extensively studied, analytical treatments of localized surface loading remain relatively scarce~\cite{Hiramatsu66,Guerrero72,Wu06,Jingu85_3D,Sato24_3D}. 
Especially, explicit closed-form expressions for the full three-dimensional stress tensor inside a finite sphere are rarely available.

From an experimental perspective, this problem is closely related to three-dimensional photoelasticity~\cite{Frocht,Coker,Yokoyama23,Yu25}. 
Reliable theoretical reference solutions for the principal stress difference inside a bounded three-dimensional body are therefore indispensable for interpreting experimental observations and validating reconstruction techniques.

In this paper, we derive an analytical solution for the elastic stress field inside a homogeneous, isotropic solid sphere subjected to a concentrated normal surface load. 
The analysis is formulated within three-dimensional linearized elastodynamics~\cite{Fung,Eringen} and employs scalar and vector potential representations together with spherical harmonic expansions. 
All expansion coefficients are determined explicitly from the traction boundary conditions, and the static solution is obtained rigorously as the long-time limit of the dynamical formulation.

A central feature of the present work is that the full stress tensor is obtained in closed form. 
This allows direct evaluation of the principal stresses and their differences throughout the interior of the sphere. 
Furthermore, by exploiting rotational symmetry, the solution derived for a load applied at the pole is generalized to arbitrary loading positions on the spherical surface. 
This provides a systematic framework for treating multiple concentrated loads through superposition.

Several studies~\cite{Sternberg52, Guerrero72} have investigated the elastic response of a sphere under concentrated loads.
These studies provide series representations of the solution; however, the resulting expressions are typically given in implicit or highly coupled forms, which makes direct evaluation and extension to dynamic problems less straightforward.
In contrast, the present study derives a new-form series solution in which all coefficients are obtained explicitly.
This feature enables efficient numerical evaluation and provides a unified framework applicable to both static and dynamic problems.

The remainder of this paper is organized as follows.
In Section~\ref{eq:setup}, we define the mechanical setup and boundary conditions.
Section~\ref{sec:elastodynamics} formulates the governing equations of linearized elastodynamics and derives the general solution using potential representations.
In Section~\ref{sec:static}, the static elastic solution is obtained as the long-time limit of the dynamical formulation, and the stress components are evaluated explicitly.
This section also presents the calculation of the principal stress difference and discusses its spatial characteristics. 
The solution is further extended to arbitrary loading positions through rotational transformations, enabling systematic construction of multi-load configurations.
Section~\ref{sec:dynamic} provides representative results of the transient (dynamical) solution and illustrates the associated wave propagation behavior.
Finally, Section~\ref{sec:conclusion} summarizes the main results and discusses their implications for three-dimensional photoelasticity and related applications.
In Appendix~\ref{sec:FEM}, the validity of the present formulation is confirmed through comparison with finite element method (FEM) simulations. 

\section{Problem formulation}\label{eq:setup}
\begin{figure}[htbp]
    \centering
    \includegraphics[width=0.75\linewidth]{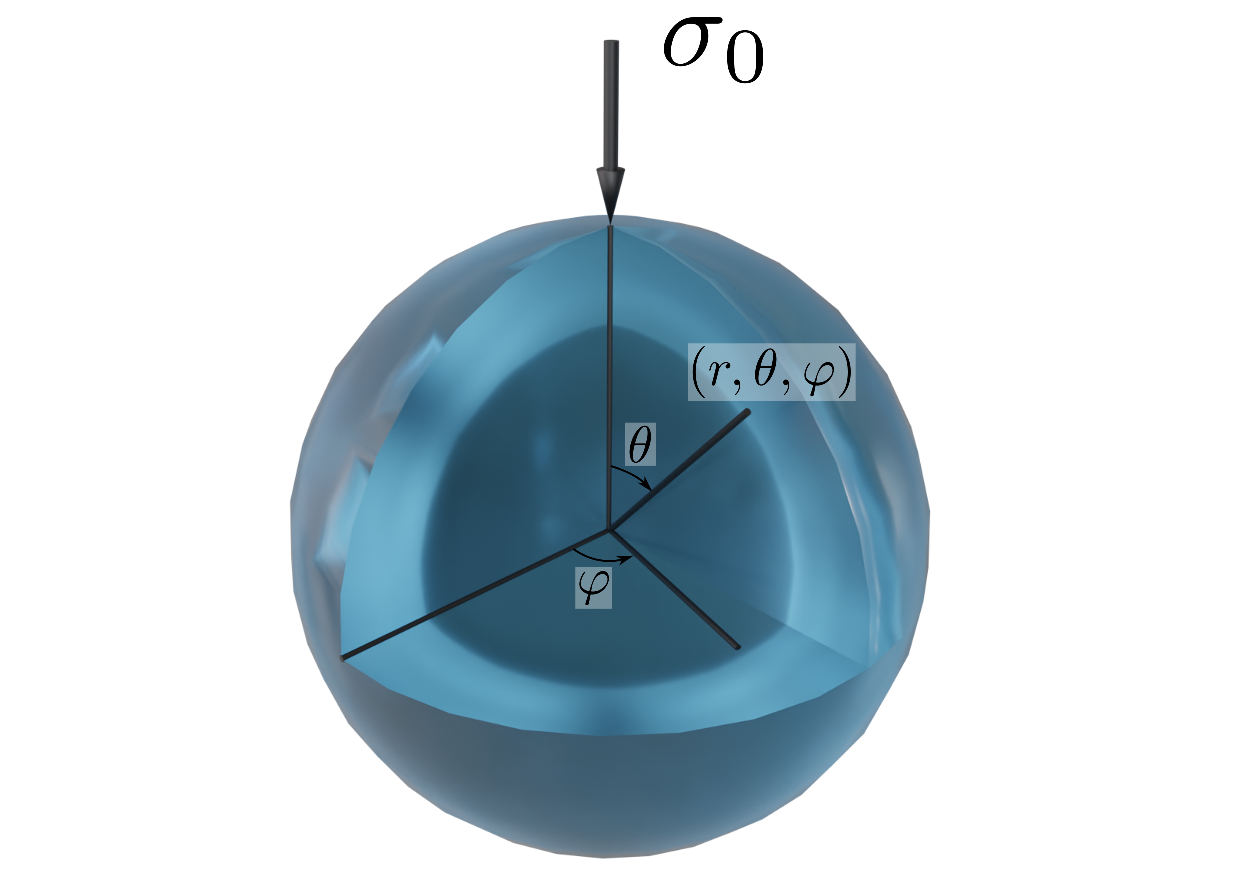}
    \caption{Schematic illustration of the system considered in this study. 
    A concentrated normal compressive stress $\sigma_0$ is applied at the north pole of a homogeneous solid elastic sphere of radius $R$.}
    \label{fig:setup}
\end{figure}
We consider a homogeneous, isotropic, solid elastic sphere of radius $R$. 
The material is characterized by the shear modulus $G$, Poisson's ratio $\nu$, and mass density $\varrho_0$. 
Our objective is to determine, within the framework of linear elasticity, the stress distribution inside the sphere when an external load is applied on its outer surface $r=R$.

Assuming the material obeys linear elasticity, the principle of superposition holds. 
Accordingly, the total stress field can be constructed by summing the stress fields generated by individual loading components. 
This property allows us to treat singular surface tractions by expanding them in an appropriate functional basis.

We focus on the configuration shown in Fig.~\ref{fig:setup}, where a concentrated normal stress of magnitude $\sigma_0$ is applied at the north pole $(x,y,z)=(0,0,R)$ of the sphere. 
The load acts in the inward normal direction and therefore corresponds to a compressive stress.
It should be noted, strictly speaking, that such a single concentrated load does not satisfy the global force balance condition and would induce a rigid-body motion of the sphere. 
In the present analysis, however, we are concerned with the internal stress distribution generated by the applied traction. 
The rigid-body translation does not affect the stress field and will therefore not be considered explicitly in what follows.

Under the assumption of axisymmetry about the $z$-axis, the boundary conditions at the outer surface $r=R$ are written as
\begin{equation}
    \left.\sigma_{rr}\right|_{r=R} 
    = -\sigma_0 \delta(\theta)\Theta(t),\quad
    \left.\sigma_{r\theta}\right|_{r=R}
    = 0,
\end{equation}
where $\delta(\theta)$ denotes the Dirac delta distribution in the polar angle $\theta$, and $\Theta(t)$ is the Heaviside step function.
For analytical convenience, the angular delta function is expanded in terms of Legendre polynomials $P_n(\cos\theta)$~\cite{Abramowitz} as
\begin{equation}
    \delta(\theta)
    = \sum_{n=0}^{\infty}
    \frac{2n+1}{2}P_n(\cos\theta).
\end{equation}
Here $P_n(x)$ denotes the Legendre polynomial of degree $n$, which satisfies the orthogonality relation
\begin{equation}
    \int_0^\pi 
    P_m(\cos\theta)\,
    P_n(\cos\theta)\,
    \sin\theta \, \mathrm{d}\theta
    =
    \frac{2}{2n+1}\delta_{mn},
\end{equation}
where $\delta_{mn}$ is the Kronecker delta.

In the following sections, we derive the stress field inside the sphere that satisfies these boundary conditions.

\section{Equations in linearized elastodynamics}\label{sec:elastodynamics}

\subsection{Governing equation}
In this section, we formulate the governing equation for the displacement and stress fields within the framework of linearized elastodynamics.

We consider the three-dimensional elastic body described in the previous section. 
Under the assumption of small deformations, the displacement field $\bm{u}(\bm{r},t)$ satisfies the Navier--Cauchy equation \cite{Timoshenko,Eringen,Fung}
\begin{equation}
    \varrho_0\frac{\partial^2}{\partial t^2}\bm{u}
    = G\left[\nabla^2 \bm{u} + \frac{1}{1-2\nu}\bm\nabla \left(\bm\nabla \cdot \bm{u}\right)\right].
    \label{eq:Navier-Cauchy}
\end{equation}
In three-dimensional elasticity, wave propagation is characterized by the longitudinal (P-wave) and transverse (S-wave) velocities~\cite{Fung,Aki,Sato24_3D},
\begin{equation}
    v_\mathrm{L}\equiv \sqrt{\frac{2(1-\nu)}{1-2\nu}\frac{G}{\varrho_0}},\quad
    v_\mathrm{T}\equiv \sqrt{\frac{G}{\varrho_0}}.
\end{equation}
For later convenience, we introduce their ratio
\begin{equation}
    \mu\equiv \frac{v_\mathrm{L}}{v_\mathrm{T}}
    = \sqrt{\frac{2(1-\nu)}{1-2\nu}}.
\end{equation}
To simplify the analysis, we nondimensionalize the variables as
\begin{align}
    \bm{\rho}\equiv \frac{\bm{r}}{R},\quad
    \tau \equiv \frac{v_\mathrm{L}t}{R},\quad
    \widetilde{\bm{u}}\equiv \frac{G}{\sigma_0}\frac{\bm{u}}{R},\quad
    \overleftrightarrow{{\widetilde{\sigma}}}
    =\frac{\overleftrightarrow{\sigma}}{\sigma_0},
\end{align}
where $\overleftrightarrow{\sigma}$ denotes the Cauchy stress tensor.

Using these dimensionless variables, Eq.~\eqref{eq:Navier-Cauchy} becomes
\begin{equation}
    \mu^2 \frac{\partial^2}{\partial \tau^2}\widetilde{\bm{u}}
    = \nabla_{\bm{\rho}}^2 \widetilde{\bm{u}} 
    + \frac{1}{1-2\nu} \bm\nabla_{\bm{\rho}} \left(\bm\nabla_{\bm{\rho}} \cdot \widetilde{\bm{u}}\right),
    \label{eq:Navier-Cauchy2}
\end{equation}
where $\bm{\nabla}_{\bm{\rho}} \equiv R \bm{\nabla}$.

\subsection{Laplace-transformed solution}\label{sec:Laplace-transformed solution}
To solve Eq.~\eqref{eq:Navier-Cauchy2}, we employ the Helmholtz decomposition of the displacement field. 
We assume that the dimensionless displacement can be expressed in terms of two scalar potentials $\varphi$ and $\chi$ as~\cite{Eringen,Jingu85_3D,Sato24_3D}
\begin{equation}
    \widetilde{\bm{u}}=\bm\nabla_{\bm{\rho}} \varphi 
    - \bm\nabla_{\bm{\rho}} \times \bm\nabla_{\bm{\rho}} \times 
    \left(\rho\chi\bm{e}_\rho\right).
    \label{eq:u_Helmholtz}
\end{equation}
where $\bm{e}_\rho$ denotes the unit vector in the radial direction.
Substituting Eq.~\eqref{eq:u_Helmholtz} into Eq.~\eqref{eq:Navier-Cauchy2}, one finds that the governing equation is satisfied provided that the scalar potentials obey the wave equations
\begin{equation}
    \nabla_{\bm{\rho}}^2 \varphi
    = \frac{\partial^2 \varphi}{\partial \tau^2},
    \quad
    \nabla_{\bm{\rho}}^2 \chi
    = \mu^2 \frac{\partial^2 \chi}{\partial \tau^2}.
    \label{eq:wave_eqs}
\end{equation}
Thus, the elastodynamic problem is reduced to solving two scalar wave equations corresponding to the longitudinal (P-wave) and transverse (S-wave) modes, respectively.

To solve Eq.~\eqref{eq:wave_eqs}, it is convenient to introduce the Laplace transforms~\cite{Abramowitz} of the scalar potentials,
\begin{equation}
    \overline{\varphi}(s)
    \equiv \int_0^\infty \varphi\, \mathrm{e}^{-s\tau}\mathrm{d}\tau, \quad
    \overline{\chi}(s)
    \equiv \int_0^\infty \chi\, \mathrm{e}^{-s\tau}\mathrm{d}\tau.
\end{equation}
We assume that the system is initially at rest for $\tau <0$, namely
\begin{equation}
    \varphi = \partial_\tau \varphi = \chi = \partial_\tau \chi = 0
    \quad \text{at } \tau=0.
\end{equation}
Under this assumption, the wave equations \eqref{eq:wave_eqs} reduce in Laplace space to
\begin{equation}
    \nabla_{\bm{\rho}}^2 \overline{\varphi} 
    = s^2\overline{\varphi},\quad
    \nabla_{\bm{\rho}}^2 \overline{\chi} 
    = \mu^2s^2\overline{\chi},
    \label{eq:wave_eqs2}
\end{equation}
which are modified Helmholtz equations.
These equations can be solved by separation of variables in spherical coordinates. 
Imposing regularity at the origin $\rho=0$, the solutions are expressed as
\begin{equation}
    \overline{\varphi}(\rho,\theta,s) 
    = \sum_{n=0}^\infty a_n i_n(\rho s)P_n(\cos\theta),\quad
    \overline{\chi} (\rho,\theta,s)
    = \sum_{n=0}^\infty b_n i_n(\mu \rho s)P_n(\cos\theta),
    \label{eq:phi_chi_sol}
\end{equation}
where $i_n(x)$ denotes the modified spherical Bessel function of the first kind~\cite{Abramowitz}. 
The modified spherical Bessel function of the second kind $k_n(x)$ is excluded because it diverges at the origin. 
The coefficients $a_n$ and $b_n$ are determined from the boundary conditions.

To determine these coefficients, we express the Laplace transforms of the displacement and stress fields as
\begin{equation}
    \overline{\bm{u}}
    \equiv \int_0^\infty \bm{u} \mathrm{e}^{-s\tau}\mathrm{d}\tau, \quad
    \overleftrightarrow{\overline{\sigma}}(s)
    \equiv \int_0^\infty \overleftrightarrow{\sigma} \mathrm{e}^{-s\tau}\mathrm{d}\tau.   
\end{equation}
Using the Helmholtz representation \eqref{eq:u_Helmholtz}, the dimensionless displacement and stress components can be written in terms of $\overline{\varphi}$ and $\overline{\chi}$ as~\cite{Jingu85_3D,Sato24_3D}
\begin{subequations}\label{eq:u_sigma_potential}
\begin{align}
    \overline{\widetilde{u}}_\rho 
    &= \frac{\partial \overline{\varphi}}{\partial \rho}
    + \frac{\cot\theta}{\rho}\frac{\partial \overline{\chi}}{\partial\theta}
    + \frac{1}{\rho}
    \frac{\partial^2 \overline{\chi}}{\partial \theta^2},\\
    \overline{\widetilde{u}}_\theta
    &= \frac{1}{\rho} \frac{\partial \overline{\varphi}}{\partial \theta}
    - \frac{\partial^2 \overline{\chi}}{\partial \rho \partial\theta}
    - \frac{1}{\rho}\frac{\partial \overline{\chi}}{\partial\theta},\\
    \frac{\overline{\widetilde{\sigma}}_{\rho\rho}}{2}
    &= \frac{\nu}{1-2\nu}s^2 \overline{\varphi}
    + \frac{\partial^2 \overline{\varphi}}{\partial \rho^2}
    - \frac{1}{\rho^2}\left(\cot\theta \frac{\partial \overline{\chi}}{\partial \theta}
    + \frac{\partial^2 \overline{\chi}}{\partial \theta^2}\right)
    + \frac{1}{\rho}\left(\cot\theta \frac{\partial^2 \overline{\chi}}{\partial \rho \partial \theta}
    + \frac{\partial^3 \overline{\chi}}{\partial \rho \partial \theta^2}\right),\\
    \frac{\overline{\widetilde{\sigma}}_{\theta\theta}}{2}
    &= \frac{\nu}{1-2\nu}s^2 \overline{\varphi}
    + \frac{1}{\rho}\frac{\partial \overline{\varphi}}{\partial \rho}
    + \frac{1}{\rho^2} \frac{\partial^2 \overline{\varphi}}{\partial \theta^2}
    + \frac{\cot\theta}{\rho^2}\frac{\partial \overline{\chi}}{\partial \theta}
    - \frac{1}{\rho}\frac{\partial^3 \overline{\chi}}{\partial \rho \partial \theta^2},\\
    \frac{\overline{\widetilde{\sigma}}_{\phi\phi}}{2}
    &= \frac{\nu}{1-2\nu}s^2 \overline{\varphi}
    + \frac{1}{\rho}\frac{\partial \overline{\varphi}}{\partial \rho}
    + \frac{\cot\theta}{\rho^2} \frac{\partial \overline{\varphi}}{\partial \theta}
    + \frac{1}{\rho^2}
    \frac{\partial^2 \overline{\chi}}{\partial \theta^2}
    - \frac{\cot\theta}{\rho}\frac{\partial^2 \overline{\chi}}{\partial \rho \partial \theta},\\
    \frac{\overline{\widetilde{\sigma}}_{\rho\theta}}{2}
    &= \frac{1}{\rho}\frac{\partial^2 \overline{\varphi}}{\partial \rho \partial \theta}
    -\frac{1}{\rho^2} \frac{\partial \overline{\varphi}}{\partial \theta}
    + \frac{1}{2}\left[
    - \frac{\partial^3 \overline{\chi}}{\partial \rho^2 \partial \theta}
    + \frac{1-\cot^2\theta}{\rho^2}\frac{\partial \overline{\chi}}{\partial \theta}
    + \frac{\cot\theta}{\rho^2}\frac{\partial^2 \overline{\chi}}{\partial \theta^2}
    + \frac{1}{\rho^2}
    \frac{\partial^3 \overline{\chi}}{\partial \theta^3}\right].
\end{align}
\end{subequations}
Substituting Eq.~\eqref{eq:phi_chi_sol} into Eq.~\eqref{eq:u_sigma_potential}, the displacement and stress fields can be expanded in terms of Legendre polynomials as
\begin{subequations}
\begin{align}
    \rho \overline{\widetilde{u}}_\rho 
    &= \sum_{n=0}^\infty \left[a_n i_n^{(1)}(\rho s)
    - n(n+1)b_ni_n(\mu\rho s)\right] P_n(\cos\theta),\\
    \rho \overline{\widetilde{u}}_\theta
    &= \sum_{n=0}^\infty \left[-a_n i_n(\rho s)
    + b_n F_{n,0}(\mu\rho s)\right]
    \sin\theta P_n^\prime(\cos\theta),\\
    \frac{\rho^2}{2}\overline{\widetilde{\sigma}}_{\rho\rho}
    &= \sum_{n=0}^\infty \left[
    a_n F_{n,1}(\rho s)- b_n n(n+1) F_{n,2}(\mu\rho s)\right] P_n(\cos\theta),
    \label{eq:bar_sigma_rr}\\
    \frac{\rho^2}{2}\overline{\widetilde{\sigma}}_{\rho\theta}
    &= \sum_{n=0}^\infty \left[-a_n F_{n,2}(\rho s) 
    + b_n F_{n,3}(\mu \rho s)\right]
    \sin\theta P_n^\prime(\cos\theta),
    \label{eq:bar_sigma_rt}\\
    \frac{\rho^2}{2}\overline{\sigma}_{\theta\theta}
    &= \sum_{n=0}^\infty \Big\{
    \left[-a_n F_{n,4}(\rho s)
    + b_n n(n+1)i_n^{(1)}(\mu \rho s)
    \right] P_n(\cos\theta)\nonumber\\
    &\hspace{4em}
    + \left[a_n i_n(\rho s)
    - b_n F_{n,0}(\mu \rho s)\right] \cos\theta P_n^\prime(\cos\theta)\Big\},\\
    \frac{\rho^2}{2}\overline{\widetilde{\sigma}}_{\phi\phi}
    &= \sum_{n=0}^\infty \Big\{
    \left[a_n F_{n,5}(\rho s)
    - b_n n(n+1)i_n(\mu\rho s)\right] P_n(\cos\theta)\nonumber\\
    &\hspace{4em}
    - \left[a_n i_n(\rho s)
    - b_n F_{n,0}(\mu\rho s)\right]
    \cos\theta P_n^\prime(\cos\theta)
    \Big\}.
\end{align}
\end{subequations}
Here we have introduced the auxiliary functions
\begin{subequations}
\begin{align}
    i_n^{(1)}(s)
    &\equiv n i_n(s) + si_{n+1}(s),\\
    F_{n,0}(s) 
    &\equiv (n+1)i_n(s) + si_{n+1}(s),\\
    F_{n,1}(s) 
    &\equiv \left[n(n-1)+\frac{1-\nu}{1-2\nu}s^2\right]i_n(s) -2si_{n+1}(s),\\
    F_{n,2}(s)
    &\equiv (n-1)i_n(s) + s i_{n+1}(s),\\
    F_{n,3}(s)
    &\equiv \left(n^2-1+\frac{s^2}{2}\right)i_n(s) - s i_{n+1}(s),\\
    F_{n,4}(s)
    &\equiv \left(n^2-\frac{\nu}{1-2\nu}s^2\right) i_n(s) - s i_{n+1}(s),\\
    F_{n,5}(s)
    &\equiv \left(n + \frac{\nu}{1-2\nu}s^2\right)i_n(s)
    + si_{n+1}(s).
\end{align}
\end{subequations}

\subsection{Determination of the coefficients}
Applying the boundary conditions at $\rho = 1$ to Eqs.~\eqref{eq:bar_sigma_rr} and \eqref{eq:bar_sigma_rt}, the coefficients $a_n$ and $b_n$ satisfy
\begin{subequations}
\begin{align}
    a_n F_{n,1}(s) 
    - b_n n(n+1) F_{n,2}(\mu s)
    &= -\frac{1}{2}\frac{2n+1}{2}\frac{1}{s},\\
    a_n F_{n,2}(s) 
    - b_n F_{n,3}(\mu s)&= 0.
\end{align}
\end{subequations}
Solving this linear system yields
\begin{align}
    a_n 
    = -\frac{1}{2}\frac{2n+1}{2}\frac{F_{n,3}(\mu s)}{s D_n(s)},\quad
    b_n 
    = -\frac{1}{2}\frac{2n+1}{2}\frac{F_{n,2}(s)}{s D_n(s)},
\end{align}
where
\begin{equation}
    D_n(s)
    \equiv F_{n,1}(s)F_{n,3}(\mu s) - n(n+1)F_{n,2}(s) F_{n,2}(\mu s).
\end{equation}
Substituting these expressions into the expansions for displacement and stress, we obtain
\begin{subequations}
\begin{align}
    \rho\overline{\widetilde{u}}_\rho 
    &= -\frac{1}{2}\sum_n \frac{2n+1}{2}
    \frac{N_{n,1}(s)}{sD_n(s)}P_n(\cos\theta),\\
    \rho\overline{\widetilde{u}}_\theta
    &= \frac{1}{2}\sum_n \frac{2n+1}{2}
    \frac{N_{n,2}(s)}{sD_n(s)}
    \sin\theta P_n^\prime(\cos\theta),\\
    \overline{\widetilde{\sigma}}_{\rho\rho}
    &= -\frac{1}{\rho^2}\sum_n \frac{2n+1}{2}
    \frac{N_{n,3}(s)}{sD_n(s)} P_n(\cos\theta),
    \label{eq:sigma_rr}\\
    \overline{\widetilde{\sigma}}_{\rho\theta}
    &= \frac{1}{\rho^2}\sum_n \frac{2n+1}{2}
    \frac{N_{n,4}(s)}{sD_n(s)}
    \sin\theta P_n^\prime(\cos\theta),
    \label{eq:sigma_rt}\\
    \overline{\widetilde{\sigma}}_{\theta\theta}
    &= \frac{1}{\rho^2}\sum_n \frac{2n+1}{2}\left[
    \frac{N_{n,5}(s)}{sD_n(s)} P_n(\cos\theta)
    - \frac{N_{n,2}(s)}{sD_n(s)} \cos\theta P_n^\prime(\cos\theta)\right],\\
    \overline{\widetilde{\sigma}}_{\phi\phi}
    &= -\frac{1}{\rho^2}\sum_n \frac{2n+1}{2}\left[
    \frac{N_{n,6}(s)}{sD_n(s)} P_n(\cos\theta)
    - \frac{N_{n,2}(s)}{sD_n(s)} \cos\theta P_n^\prime(\cos\theta)\right].
\end{align}
\end{subequations}
Here the auxiliary functions are defined as
\begin{subequations}
\begin{align}
    N_{n,1}(s)
    &\equiv i_n^{(1)}(\rho s)F_{n,3}(\mu s)
    - n(n+1)F_{n,2}(s)i_n(\mu\rho s),\\
    N_{n,2}(s)
    &\equiv i_n(\rho s)F_{n,3}(\mu s) 
    - F_{n,2}(s)F_{n,0}(\mu\rho s),\\
    N_{n,3}(s)
    &\equiv F_{n,1}(\rho s)F_{n,3}(\mu s) 
    - n(n+1) F_{n,2}(s)F_{n,2}(\mu\rho s),\\
    N_{n,4}(s)
    &\equiv F_{n,2}(\rho s)F_{n,3}(\mu s) 
    - F_{n,2}(s)F_{n,3}(\mu \rho s),\\
    N_{n,5}(s)
    &\equiv F_{n,4}(\rho s)F_{n,3}(\mu s)
    - n(n+1)F_{n,2}(s)i_n^{(1)}(\mu \rho s),\\
    N_{n,6}(s)
    &\equiv F_{n,5}(\rho s)F_{n,3}(\mu s)
    - n(n+1)F_{n,2}(s)i_n(\mu \rho s).
\end{align}
\end{subequations}

Finally, performing the inverse Laplace transform
$\mathcal{L}^{-1}$, the time-dependent displacement and stress fields are obtained as
\begin{subequations}\label{eq:u_sigma_time}
\begin{align}
    \rho \widetilde{u}_\rho 
    &= -\sum_n \frac{2n+1}{4}
    \mathcal{L}^{-1}\left[\frac{N_{n,1}(s)}{sD_n(s)}\right]P_n(\cos\theta),\\
    \rho \widetilde{u}_\theta
    &= \sum_n \frac{2n+1}{4}
    \mathcal{L}^{-1}\left[\frac{N_{n,2}(s)}{sD_n(s)}\right]
    \sin\theta P_n^\prime(\cos\theta),\\
    \widetilde{\sigma}_{\rho\rho}
    &= -\frac{1}{\rho^2}\sum_n \frac{2n+1}{2}
    \mathcal{L}^{-1}\left[\frac{N_{n,3}(s)}{sD_n(s)}\right] P_n(\cos\theta),
    \label{eq:sigma_rr_inv}\\
    \widetilde{\sigma}_{\rho\theta}
    &= \frac{1}{\rho^2}\sum_n \frac{2n+1}{2}
    \mathcal{L}^{-1}\left[\frac{N_{n,4}(s)}{sD_n(s)}\right]
    \sin\theta P_n^\prime(\cos\theta),\\
    \widetilde{\sigma}_{\theta\theta}
    &= \frac{1}{\rho^2}\sum_n \frac{2n+1}{2}\left\{
    \mathcal{L}^{-1}\left[\frac{N_{n,5}(s)}{sD_n(s)}\right] P_n(\cos\theta)\right.\nonumber\\
    &\hspace{9em}\left.
    - \mathcal{L}^{-1}\left[\frac{N_{n,2}(s)}{sD_n(s)}\right] \cos\theta P_n^\prime(\cos\theta)\right\},\\
    \widetilde{\sigma}_{\phi\phi}
    &= -\frac{1}{\rho^2}\sum_n \frac{2n+1}{2}\left\{
    \mathcal{L}^{-1}\left[\frac{N_{n,6}(s)}{sD_n(s)}\right] P_n(\cos\theta)\right.\nonumber\\
    &\hspace{9em}\left.
    - \mathcal{L}^{-1}\left[\frac{N_{n,2}(s)}{sD_n(s)}\right] \cos\theta P_n^\prime(\cos\theta)\right\}.
\end{align}
\end{subequations}

In general, it is difficult to evaluate the inverse Laplace transform analytically. 
In practical computations, numerical techniques such as fast Fourier transform–based inversion methods are commonly employed~\cite{Press}. 
Alternatively, one may evaluate the inverse transform by deforming the integration contour in the complex plane so as to form a closed contour and then applying the residue theorem~\cite{Jingu85_3D,Sato24_3D,Jingu85_2D,Sato24_2D}. 
In this approach, the integral is expressed as the sum of contributions from the poles enclosed by the contour.
In general, infinitely many poles are present. 
However, the contribution from poles located farther from the origin in the complex plane typically decreases in magnitude. 
Therefore, by summing a sufficiently large but finite number of pole contributions, one can approximate the displacement and stress fields at an arbitrary time with controllable accuracy.
Physically, the resulting time-dependent solution describes the propagation of stress waves generated by the surface load. 
In particular, compressional (P) waves and shear (S) waves propagate through the interior of the sphere~\cite{Aki,Fung,Sato24_3D,Sato24_2D}.

\section{Static solution}\label{sec:static}
In this section, we present the static solution.
First, in Sec.~\ref{sec:north}, we show the results for the case where the load is applied at the north pole. 
Next, in Sec.~\ref{sec:arbitrary_point}, we describe how to compute the stress distribution when the load is applied at an arbitrary point on the surface. 
Finally, in Sec.~\ref{sec:several_loads}, we present examples in which multiple loads act simultaneously.

\subsection{Load applied at the north pole}\label{sec:north}
The static solution is obtained in the limit $t \to \infty$~\cite{Sato24_3D,Sato24_2D,Okamura25}.
Using the final value theorem of the Laplace transform, we have
\begin{equation}
    \lim_{t\to \infty}\mathcal{L}^{-1}\left[\frac{N_{n,i}(s)}{sD_n(s)}\right]
    = \lim_{s\to 0} s \frac{N_{n,i}(s)}{sD_n(s)}
    = \lim_{s\to 0} \frac{N_{n,i}(s)}{D_n(s)}.
\end{equation}
Applying this result yields
\begin{subequations}\label{eq:u_sigma_steady2}
\begin{align}
    \widetilde{u}_\rho^{(\mathrm{st})}
    &= -\sum_{n\neq 1} 
    \frac{2n+1}{8(n-1)}\frac{n\mathcal{N}_{n,1}^{(\mathrm{st})}-\mathcal{N}_{n,2}^{(\mathrm{st})}\rho^2}{\mathcal{D}_n^{(\mathrm{st})}}\rho^{n-1}
    P_n(\cos\theta),\\
    \widetilde{u}_\theta^{(\mathrm{st})}
    &= \sum_{n\neq 1}
    \frac{2n+1}{8(n-1)}\frac{\mathcal{N}_{n,1}^{(\mathrm{st})}-\mathcal{N}_{n,3}^{(\mathrm{st})}\rho^2}{\mathcal{D}_n^{(\mathrm{st})}}\rho^{n-1}
    \sin\theta P_n^\prime(\cos\theta),\\
    \widetilde{\sigma}_{\rho\rho}^{(\mathrm{st})}
    &= -\sum_{n\neq 1} 
    \frac{2n+1}{4}
    \frac{n\mathcal{N}_{n,1}^{(\mathrm{st})}-\mathcal{N}_{n,4}^{(\mathrm{st})}\rho^2}{\mathcal{D}_n^{(\mathrm{st})}}\rho^{n-2}
    P_n(\cos\theta),\\
    \widetilde{\sigma}_{\rho\theta}^{(\mathrm{st})}
    &= \sum_{n\neq 1} 
    \frac{2n+1}{4}
    \frac{\mathcal{N}_{n,1}^{(\mathrm{st})}(1-\rho^2)}{\mathcal{D}_n^{(\mathrm{st})}}\rho^{n-2}
    \sin\theta P_n^\prime(\cos\theta),\\
    \widetilde{\sigma}_{\theta\theta}^{(\mathrm{st})}
    &= \sum_{n\neq 1} 
    \frac{2n+1}{4(n-1)}\left\{
    \frac{n^2\mathcal{N}_{n,1}^{(\mathrm{st})}-\mathcal{N}_{n,5}^{(\mathrm{st})}\rho^2}{\mathcal{D}_n^{(\mathrm{st})}}\rho^{n-2}
    P_n(\cos\theta)\right.\nonumber\\
    &\hspace{8em}\left.
    - \frac{\mathcal{N}_{n,1}^{(\mathrm{st})}-\mathcal{N}_{n,3}^{(\mathrm{st})}\rho^2}{\mathcal{D}_n^{(\mathrm{st})}}\rho^{n-2}
    \cos\theta P_n^\prime(\cos\theta)\right\},\\
    \widetilde{\sigma}_{\phi\phi}^{(\mathrm{st})}
    &= -\sum_{n\neq 1} 
    \frac{2n+1}{4(n-1)}\left\{
    \frac{n\mathcal{N}_{n,1}^{(\mathrm{st})} - \mathcal{N}_{n,6}^{(\mathrm{st})}\rho^2}{\mathcal{D}_n^{(\mathrm{st})}}\rho^{n-2}
    P_n(\cos\theta)\right.\nonumber\\
    &\hspace{8em}\left.
    - \frac{\mathcal{N}_{n,1}^{(\mathrm{st})}-\mathcal{N}_{n,3}^{(\mathrm{st})}\rho^2}{\mathcal{D}_n^{(\mathrm{st})}}\rho^{n-2}
    \cos\theta P_n^\prime(\cos\theta)\right\}.
\end{align}
\end{subequations}
Here the auxiliary functions are defined as
\begin{subequations}
\begin{align}
    \mathcal{D}_n^{(\mathrm{st})}
    &\equiv n^2+n+1+(2n+1)\nu,\\
    \mathcal{N}_{n,1}^{(\mathrm{st})}
    &\equiv n^2+2n-1+2\nu,\quad
    \mathcal{N}_{n,2}^{(\mathrm{st})}
    \equiv (n^2-1)(n-2+4\nu),\\
    \mathcal{N}_{n,3}^{(\mathrm{st})}
    &\equiv (n-1)(n+5-4\nu),\quad
    \mathcal{N}_{n,4}^{(\mathrm{st})}
    \equiv (n+1)[(n-2)(n+1)-2\nu],\\
    \mathcal{N}_{n,5}^{(\mathrm{st})}
    &\equiv (n^2-1)(n^2+4n+2+2\nu),\quad
    \mathcal{N}_{n,6}^{(\mathrm{st})}
    \equiv (n^2-1)\left[n-2-2(2n+1)\nu\right].
\end{align}
\end{subequations}
Hereafter in this section, we denote $\widetilde{\sigma}_{\alpha\beta}^{(\mathrm{st})}$ simply as $\widetilde{\sigma}_{\alpha\beta}$ for simplicity.

The $n=1$ mode has been excluded from the above expressions. 
This mode corresponds to a rigid-body translation of the sphere. 
Indeed, the total force acting on the sphere is determined solely by the $n=1$ component of the boundary traction, owing to the orthogonality of the Legendre polynomials. 
Therefore, the $n=1$ term does not contribute to internal elastic deformation but instead represents a global translational motion. 
Since the present study focuses on the internal stress distribution, this rigid-body mode can be removed without loss of generality.

\begin{figure}[htbp]
    \centering
    \includegraphics[width=0.75\linewidth]{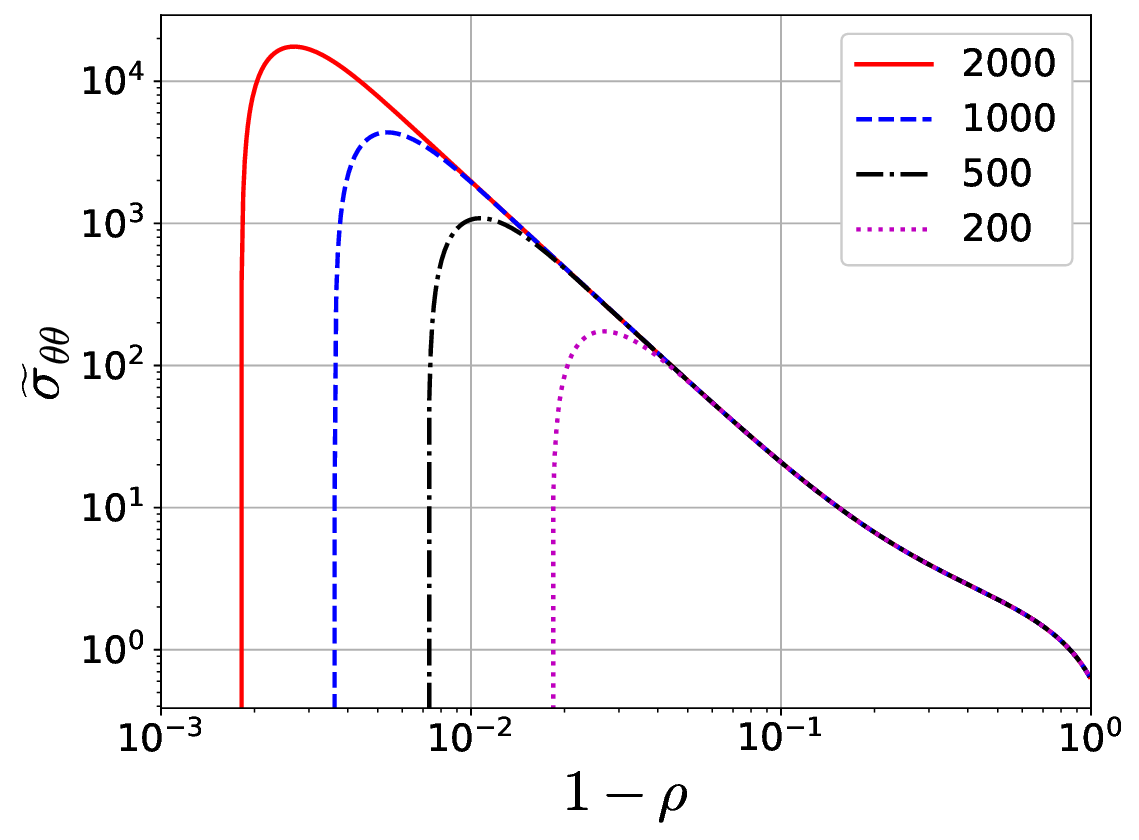}
    \caption{Profile of $\widetilde{\sigma}_{\theta\theta}$ for $N_{\mathrm{tr}} = 200$ (dotted line), $500$ (dash-dotted line), $1000$ (dashed line), and $2000$ (solid line) at $\nu = 0.3$. 
    The $n=1$ mode is excluded.}
    \label{fig:convergence}
\end{figure}

In practical computations, the infinite series must be truncated at a finite order.
Here, we denote the truncation order by $N_{\mathrm{tr}}$ and examine the convergence behavior of Eq.~\eqref{eq:u_sigma_steady2}.
Figure~\ref{fig:convergence} shows the profile of $\widetilde{\sigma}_{\theta\theta}$ along the loading axis ($\theta=0$) for several values of $N_\mathrm{tr}$.
The behavior of the other stress components on the loading axis will be discussed later.
In regions sufficiently far from the loading point, the solution is already well converged even for $N_\mathrm{tr} = 200$.
However, as the loading point is approached, truncation errors become significant.
In particular, although the exact solution exhibits a divergent behavior toward $+\infty$, the truncated series instead shows an artificial divergence toward $-\infty$.
This discrepancy is an unavoidable consequence of truncating the infinite series.
In practical applications, however, this issue can be mitigated by ensuring that such errors are confined to a region smaller than the characteristic mesh size of the discretization.
In the following, we therefore adopt $N_{\mathrm{tr}} = 2000$, which provides sufficient accuracy over the region of interest.

Let us visualize the result.
We begin with the case where the concentrated load is applied at the north pole. 
In photoelastic experiments~\cite{Frocht,Coker,Yokoyama23,Yu25}, stresses are observed through interference fringes, which are determined by the principal stress difference. 
The principal stresses are obtained as the eigenvalues of the stress tensor
\begin{equation}
    \begin{pmatrix}
    \widetilde{\sigma}_{\rho\rho} & \widetilde{\sigma}_{\rho\theta} & 0\\
    \widetilde{\sigma}_{\rho\theta} & \widetilde{\sigma}_{\theta\theta} & 0\\
    0 & 0 & \widetilde{\sigma}_{\phi\phi}
    \end{pmatrix}.
    \label{eq:sigma_matrix}
\end{equation}
Therefore, using the stress components derived in the previous section, we compute the eigenvalues and take the difference between the maximum and minimum values to obtain the principal stress difference.

From Eq.~\eqref{eq:sigma_matrix}, one immediately notices that $\widetilde{\sigma}_{\phi\phi}$ is decoupled from the other components. 
Indeed,
\begin{align}
    \det \begin{pmatrix}
    \lambda - \widetilde{\sigma}_{\rho\rho} & -\widetilde{\sigma}_{\rho\theta} & 0\\
    -\widetilde{\sigma}_{\rho\theta} & \lambda - \widetilde{\sigma}_{\theta\theta} & 0\\
    0 & 0 & \lambda - \widetilde{\sigma}_{\phi\phi}
    \end{pmatrix}
    = \left(\lambda - \widetilde{\sigma}_{\phi\phi}\right)
    \det \begin{pmatrix}
    \lambda - \widetilde{\sigma}_{\rho\rho} & -\widetilde{\sigma}_{\rho\theta}\\
    -\widetilde{\sigma}_{\rho\theta} & \lambda - \widetilde{\sigma}_{\theta\theta}
    \end{pmatrix},
\end{align}
so that the three eigenvalues are given by
\begin{equation}
    \lambda 
    = \widetilde{\sigma}_{\phi\phi},\quad
    \frac{\widetilde{\sigma}_{\rho\rho}+\widetilde{\sigma}_{\theta\theta}}{2}
    \pm \sqrt{\left(\frac{\widetilde{\sigma}_{\rho\rho}-\widetilde{\sigma}_{\theta\theta}}{2}\right)^2 + \left(\widetilde{\sigma}_{\rho\theta}\right)^2}.
\end{equation}

\begin{figure}[htbp]
    \centering
    \includegraphics[width=0.75\linewidth]{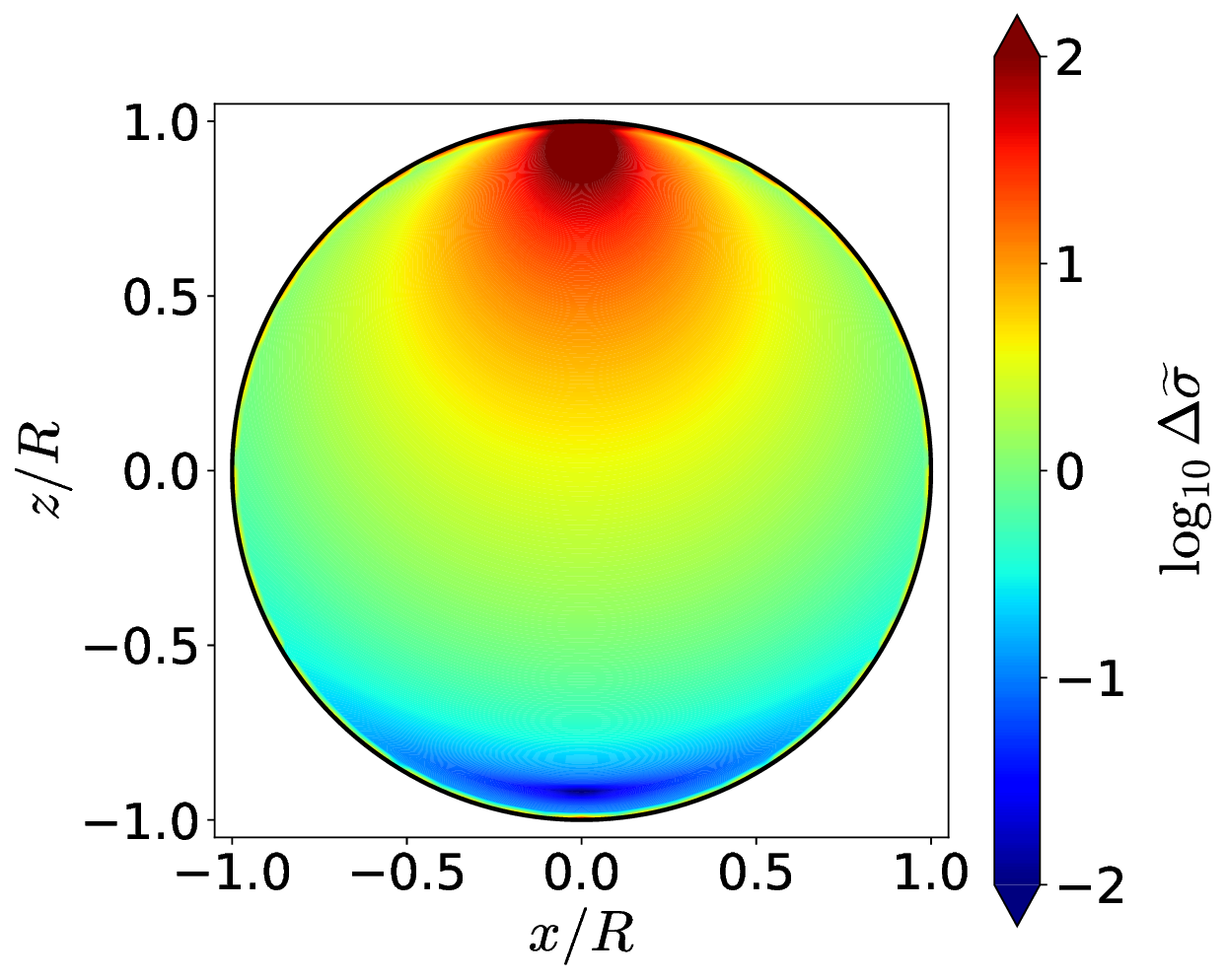}
    \caption{Principal stress difference for $\nu=0.3$. 
    The $n=1$ mode has been excluded.}
    \label{fig:del_sigma_N}
\end{figure}

Figure~\ref{fig:del_sigma_N} shows the principal stress difference for $\nu=0.3$. 
The summation over $n$ is taken up to $n=2000$, and the $n=1$ mode has been excluded as discussed previously.
The principal stress difference is largest in the vicinity of the north pole, where the load is applied. 
From this region, it decreases almost concentrically toward the interior of the sphere. 
This behavior reflects the strong stress concentration near the loading point and the subsequent three-dimensional redistribution of stress inside the elastic body.

\begin{figure}[htbp]
    \centering
    \includegraphics[width=0.75\linewidth]{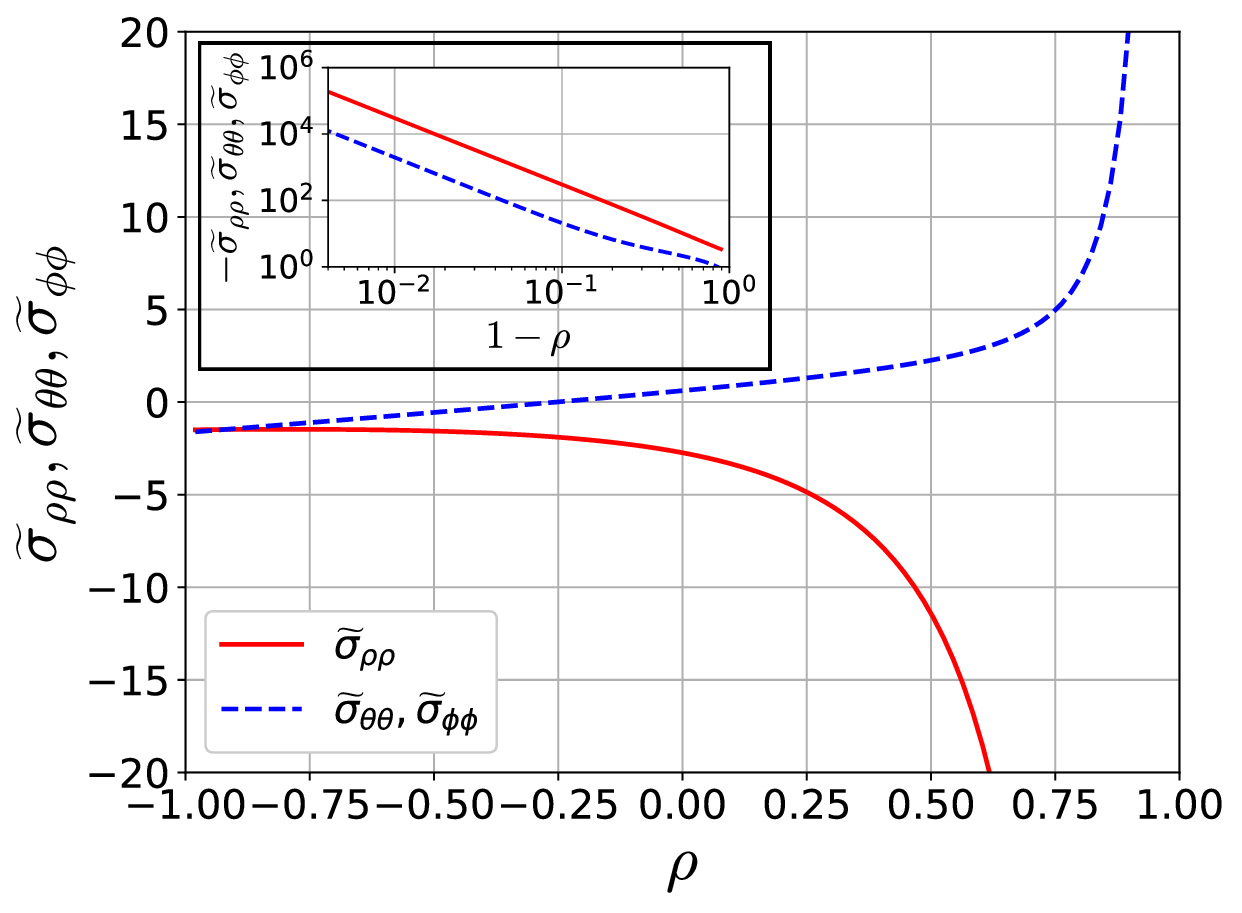}
    \caption{Stress components along the loading axis for $\nu=0.3$. 
    The inset shows the near-surface behavior ($\rho \lesssim 1$). 
    The $n=1$ mode has been excluded.}
    \label{fig:loading_line}
\end{figure}

We next examine the stress components along the loading axis. 
On the loading axis, $\theta=0$ or $\pi$, and by symmetry $\widetilde{\sigma}_{\rho\theta}=0$. 
Furthermore, $\widetilde{\sigma}_{\theta\theta}=\widetilde{\sigma}_{\phi\phi}$ holds along this axis.
Figure~\ref{fig:loading_line} shows the profiles of $\widetilde{\sigma}_{\rho\rho}$ and $\widetilde{\sigma}_{\theta\theta}(=\widetilde{\sigma}_{\phi\phi})$ along the loading axis. 
As $\rho \to 1$, $\widetilde{\sigma}_{\rho\rho}$ becomes negative, indicating compressive stress near the loading point that penetrates into the interior. 
In contrast, $\widetilde{\sigma}_{\theta\theta}$ takes positive values. 
This reflects the lateral expansion induced by the radial compression.
An important difference from the two-dimensional case should be emphasized. 
In two-dimensional elasticity, particularly under opposing concentrated loads, the tensile stress along the loading axis remains constant. 
However, in three dimensions, stress can spread in the azimuthal direction, and therefore the tensile component does not remain constant but instead exhibits a divergent tendency near the surface.
The inset of Fig.~\ref{fig:loading_line} shows the behavior for $\rho \lesssim 1$. 
Both stress components exhibit a divergence proportional to $(1-\rho)^{-2}$, which is a notable feature of the three-dimensional stress concentration near a point load.

\subsection{Extension to an arbitrary loading point}\label{sec:arbitrary_point}
In the previous subsection, we considered the case where a concentrated load acts at the north pole $(r=R,\,\theta=0)$. 
Here we extend the formulation to the case where the load is applied at an arbitrary point $(R,\theta_0,\phi_0)$ on the surface.

We denote the stress tensor \eqref{eq:sigma_matrix} obtained for the north-pole loading (hereafter referred to as the \textit{north-pole solution}) as $\overleftrightarrow{\widetilde{\sigma}^{\mathrm{N}}}(\rho, \theta)$.
Let $(r,\theta,\phi)$ be the field point at which the stress is evaluated. 
The angle $\gamma$ between the loading point $(R,\theta_0,\phi_0)$ and the field point $(r,\theta,\phi)$ satisfies
\begin{equation}
    \cos\gamma
    = 
    \begin{pmatrix}
        \sin\theta \cos\phi\\
        \sin\theta \sin\phi\\
        \cos\theta
    \end{pmatrix}
    \cdot 
    \begin{pmatrix}
        \sin\theta_0 \cos\phi_0\\
        \sin\theta_0 \sin\phi_0\\
        \cos\theta_0
    \end{pmatrix}
    = \cos\theta \cos\theta_0
    + \sin\theta \sin\theta_0 \cos(\phi-\phi_0).
\end{equation}
The stress tensor for a load applied at $(R,\theta_0,\phi_0)$ is obtained by appropriately rotating the north-pole solution. 
It can be written as
\begin{align}
    &\overleftrightarrow{\sigma}(\rho, \theta, \phi)\nonumber\\
    &= \overleftrightarrow{A}^\mathrm{T}(\theta, \phi)
    \overleftrightarrow{Q}(\theta_0, \phi_0)
    \overleftrightarrow{A}(\gamma, 0)
    \overleftrightarrow{\widetilde{\sigma}^{\mathrm{N}}}(\rho, \gamma)
    \overleftrightarrow{A}^\mathrm{T}(\gamma, 0)
    \overleftrightarrow{Q}^\mathrm{T}(\theta_0, \phi_0)
    \overleftrightarrow{A}(\theta, \phi),
    \label{eq:sigma_general}
\end{align}
where
\begin{equation}
    \overleftrightarrow{A}(\theta,\phi)
    =
    \begin{pmatrix}
    \sin\theta\cos\phi & \cos\theta\cos\phi & -\sin\phi\\
    \sin\theta\sin\phi & \cos\theta\sin\phi & \cos\phi\\
    \cos\theta & -\sin\theta & 0
    \end{pmatrix},
\end{equation}
is the transformation matrix from Cartesian to spherical bases, and
\begin{align}
    \overleftrightarrow{Q}(\theta,\phi)
    \equiv
    \overleftrightarrow{R_z}(\phi)
    \overleftrightarrow{R_y}(\theta)
    &\equiv
    \begin{pmatrix}
    \cos\phi & -\sin\phi & 0\\
    \sin\phi & \cos\phi & 0\\
    0 & 0 & 1
    \end{pmatrix}
    \begin{pmatrix}
    \cos\theta & 0 & \sin\theta\\
    0 & 1 & 0\\
    -\sin\theta & 0 & \cos\theta
    \end{pmatrix}\nonumber\\
    &= 
    \begin{pmatrix}
    \cos\theta \cos\phi & -\sin\phi & \sin\theta \cos\phi\\
    \cos\theta \sin\phi & \cos\phi & \sin\theta\sin\phi\\
    -\sin\theta & 0 & \cos\theta
    \end{pmatrix},
\end{align}
is the rotation matrix.

\begin{figure}[htbp]
    \centering
    \includegraphics[width=0.75\linewidth]{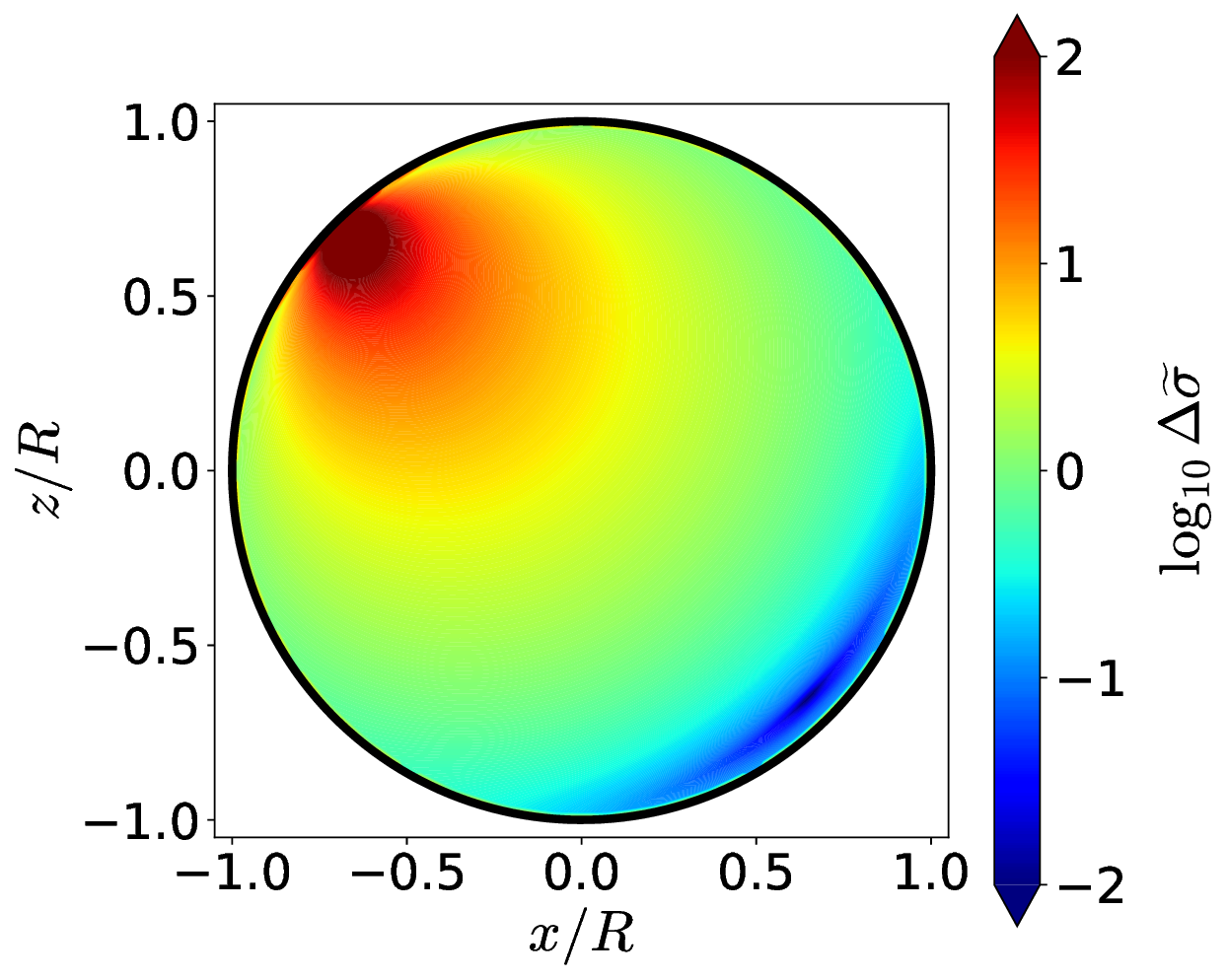}
    \caption{Principal stress difference for $\nu=0.3$ with the load applied at $(\theta_0=\pi/4,\ \phi_0=0)$. 
    The $n=1$ mode has been excluded.}
    \label{fig:del_sigma}
\end{figure}

Figure~\ref{fig:del_sigma} shows the principal stress difference for the case $\theta_0=\pi/4$ and $\phi_0=0$ with $\nu=0.3$. 
Without loss of generality, we choose the coordinate system such that $\phi_0=0$, owing to the rotational symmetry of the sphere.

Although the resulting stress distribution appears to be a simple rotation of the north-pole case, it is obtained by explicitly evaluating Eq.~\eqref{eq:sigma_general}, which consistently transforms both the tensor components and the local basis.

\subsection{Stress distribution under multiple loads}\label{sec:several_loads}
By superposing the stress fields derived above, we can compute the stress distribution under multiple concentrated loads.

\begin{figure}[htbp]
    \centering
    \includegraphics[width=0.75\linewidth]{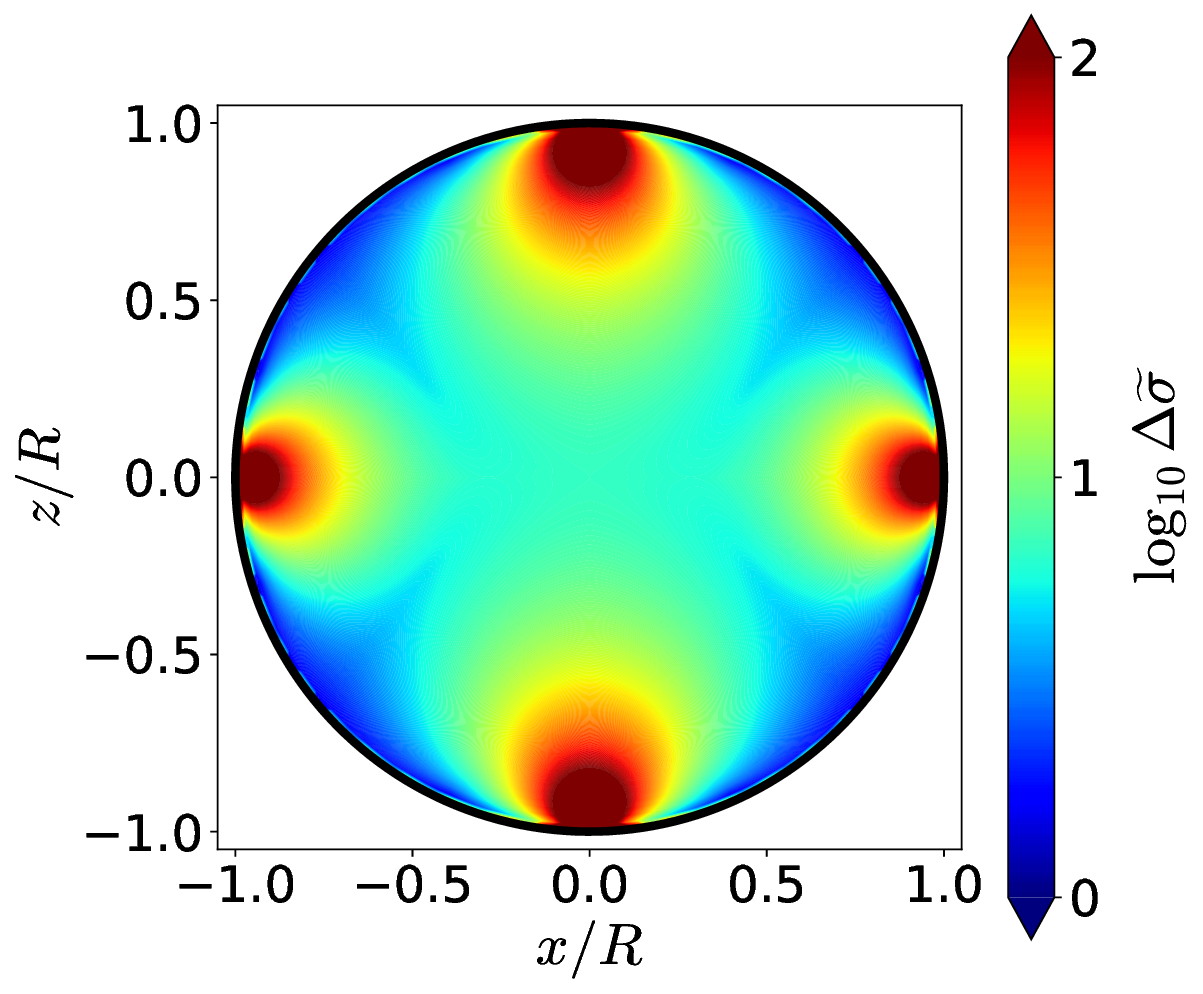}
    \caption{Principal stress difference for $\nu=0.3$ under four concentrated loads: loads of magnitude $\sigma_0$ applied at the north pole ($\theta=0$) and the south pole ($\theta=\pi$), and loads of magnitude $\sigma_0/2$ applied at $(\theta, \phi)=(\pi/2, 0)$ and $(\pi/2, \pi)$.}
    \label{fig:del_sigma_several}
\end{figure}

Figure~\ref{fig:del_sigma_several} shows the stress distribution when four loads act simultaneously. 
In this configuration, the four forces are mutually balanced, and therefore the elastic sphere can remain in static equilibrium. 
Consequently, the $n=1$ mode does not need to be removed in this case.
Due to the superposition of multiple loads, regions appear inside the sphere where stresses are amplified by constructive interaction, while in other regions the stresses are reduced because of partial cancellation. 
Although this configuration serves merely as an illustrative example, the same procedure can be applied to arbitrary loading conditions. 
Thus, by specifying a desired set of surface loads, one can systematically determine where stress concentrations occur within the sphere.

To further assess the validity of the present results, a comparison with those obtained using the finite element method (FEM) is also performed. 
Details are provided in Appendix~\ref{sec:FEM}.

\section{Dynamic solutions}\label{sec:dynamic}
In this section, we discuss the dynamic solution.
As mentioned at the end of Section \ref{sec:elastodynamics}, both P- and S-waves emerge in the elastodynamic response.
As in the case of the static solution, the formulation can be extended to arbitrary loading positions by applying an appropriate rotational transformation.
For simplicity, we therefore consider the case where the load is applied at the north pole.

To obtain the dynamic solution, it is necessary to evaluate the inverse Laplace transform in Eq.~\eqref{eq:u_sigma_time}.
Here, we focus on Eq.~\eqref{eq:sigma_rr_inv}, noting that the other quantities can be computed in an analogous manner.
The inverse Laplace transform of $N_{n,3}(s)/[sD_n(s)]$ is given by
\begin{equation}
    \mathcal{L}^{-1}\left[\frac{N_{n,3}(s)}{sD_n(s)}\right]
    = \frac{1}{2\pi \mathrm{i}}
    \int_{\gamma-\mathrm{i}\infty}^{\gamma+\mathrm{i}\infty}
    \frac{N_{n,3}(s)}{sD_n(s)}
    \mathrm{e}^{s\tau}\,\mathrm{d}s,
    \label{eq:sigma_rr_inv_integral}
\end{equation}
where $\gamma (>0)$ must be chosen such that it is greater than the real part of all singularities.

To evaluate this integral either numerically or analytically, methods such as fast Fourier transform–based inversion or contour deformation combined with the residue theorem are commonly employed.
In this study, we adopt the latter approach in order to obtain analytical expressions.

First, all poles of the integrand $N_{n,3}(s)/[sD_n(s)]$ lie on the imaginary axis and are symmetric with respect to the real axis.
Note that $s=0$ is also a pole, which corresponds to the static solution.
Accordingly, we denote the poles as $s=0$ and $s=\pm \mathrm{i}s_{n,m}$ $(0<s_{n,1}<s_{n,2}<\cdots)$.
In the limit $|s|\to\infty$, the first and second terms of $N_{n,3}(s)$ and the denominator $D_n(s)$ behave as
\begin{subequations}
\begin{align}
    &F_{n,1}(\rho s)F_{n,3}(\mu s)
    \sim \mathrm{e}^{(\rho+\mu)s},\quad
    n(n+1)F_{n,2}(s)F_{n,2}(\mu\rho s)
    \sim \mathrm{e}^{(1+\mu\rho)s},\\
    &D_n(s)\sim\mathrm{e}^{(1+\mu)s}.
\end{align}
\end{subequations}
Therefore,
\begin{subequations}\label{eq:asymptotic}
\begin{align}
    \frac{F_{n,1}(\rho s)F_{n,3}(\mu s)}{sD_n(s)}
    \mathrm{e}^{s\tau}
    &\sim \mathrm{e}^{[\tau-(1-\rho)]s},\\
    \frac{n(n+1)F_{n,2}(s)F_{n,2}(\mu\rho s)}{sD_n(s)}
    \mathrm{e}^{s\tau}
    &\sim \mathrm{e}^{[\tau-\mu(1-\rho)]s}.
\end{align}
\end{subequations}
Based on these asymptotic behaviors, we evaluate the integral by appropriately closing the contour in different regions and applying the residue theorem.

\subsection{Case for $0\le \tau<1-\rho$}
We first consider the case $0 \le \tau < 1 - \rho$.
From Eq.~\eqref{eq:asymptotic}, the integral in Eq.~\eqref{eq:sigma_rr_inv_integral} can be rewritten by adding a path in the complex plane that runs from $\mathrm{i}\infty$ to $-\mathrm{i}\infty$ through the region where $\Re(s) > 0$ and $|s| \gg 1$, thereby forming a closed contour.
Then, Eq.~\eqref{eq:sigma_rr_inv_integral} becomes
\begin{equation}
    \frac{1}{2\pi \mathrm{i}}
    \oint_C
    \frac{N_{n,3}(s)}{sD_n(s)}
    \mathrm{e}^{s\tau}\mathrm{d}s,
\end{equation}
where $C$ denotes the closed contour constructed from the original Bromwich path and the added arc.

Since all poles of the integrand lie on the imaginary axis, no poles are enclosed within the contour $C$.
Therefore, the integrand is analytic inside $C$, and by the residue theorem the integral vanishes.
Hence,
\begin{equation}
    \widetilde{\sigma}_{\rho\rho}(\rho, \theta, \tau)=0.
\end{equation}
By the same argument, all components of the displacement and stress also vanish:
\begin{equation}
    \widetilde{u}_\alpha(\rho, \theta, \tau)
    = \widetilde{\sigma}_{\alpha\beta}(\rho, \theta, \tau)
    = 0.
\end{equation}
This result is physically natural, since neither the P-wave nor the S-wave, propagating from the loading point with dimensionless velocities $1$ and $1/\mu$, respectively, has reached the observation point at this time.

\subsection{Case for $1-\rho\le \tau<\mu(1-\rho)$}
Next, we consider the case $1 - \rho \le \tau < \mu(1 - \rho)$.
From Eq.~\eqref{eq:asymptotic}, the first and second terms can be treated by adding integration paths in the complex plane that pass through regions where the real part of $s$ is negative and positive, respectively, with $|s| \gg 1$, without affecting the value of the integral.
Upon closing the contour in this manner, the contribution from the second term vanishes by the same argument as in the case $0 \le \tau < 1 - \rho$.
In contrast, the first term generally yields a nonzero contribution, since the corresponding contour encloses poles.
The condition $1 - \rho \le \tau < \mu(1 - \rho)$ indicates that only the P-wave has reached the observation point, while the S-wave has not yet arrived.

To evaluate the P-wave contribution, it is convenient to apply the residue theorem:
\begin{align}
    \widetilde{\sigma}_{\rho\rho}(\rho, \theta, \tau)
    &= -\sum_{n\neq 1}
    \frac{2n+1}{2}
    \left(\mathrm{Res}_{s=0} + 2\Re \sum_{m=1}^\infty \mathrm{Res}_{s=\mathrm{i}s_{n,m}}\right)
    \frac{F_{n,1}(\rho s)F_{n,3}(\mu s)}{\rho^2 sD_n(s)}
    \mathrm{e}^{s\tau}\nonumber\\
    &\hspace{1em}\times 
    P_n(\cos\theta),
\end{align}
and the same procedure applies to the other field quantities.
Here, $s_{n,m}$ are the positive real roots satisfying
\begin{equation}
    \Delta_n(s)\equiv D_n(\mathrm{i}s)
    = f_{n,1}(s)f_{n,3}(\mu s) -n(n+1)f_{n,2}(s)f_{n,2}(\mu s)
    = 0,
\end{equation}
with $0<s_{n,1}<s_{n,2}<\cdots$.
The functions $f_{n,i}(s)$ are defined as
\begin{subequations}
\begin{align}
    f_{n,1}(s)
    &\equiv \mathrm{i}^n F_{n,1}(\mathrm{i}s)
    = \left[n(n-1)-\frac{1-\nu}{1-2\nu}s^2\right]j_n(s)
    + 2s j_{n+1}(s),\\
    f_{n,2}(s)
    &\equiv \mathrm{i}^n F_{n,2}(\mathrm{i}s)
    = (n-1)j_n(s)-sj_{n+1}(s),\\
    f_{n,3}(s)
    &\equiv \mathrm{i}^n F_{n,3}(\mathrm{i}s)
    = \left(n^2-1-\frac{s^2}{2}\right)j_n(s) + sj_{n+1}(s),
\end{align}
\end{subequations}
Since all poles except $s=0$ are simple, the evaluation yields
\begin{equation}
    \widetilde{\sigma}_{\rho\rho}(\rho, \theta, \tau)
    = \widetilde{\sigma}_{\rho\rho}^{(\mathrm{tr})}(\rho, \theta, \tau)
    + \sum_{n\neq 1} \sum_{m=1}^\infty
    \widetilde{\sigma}_{\rho\rho}^{(\mathrm{P},n)}(\rho,\theta,s_{n,m})
    \cos(s_{n,m}\tau).
\end{equation}
Here,
\begin{subequations}\label{eq:sigma_rr_case2}
\begin{align}
    \widetilde{\sigma}_{\rho\rho}^{(\mathrm{tr})}(\rho,\theta,\tau)
    &\equiv -\sum_{n\neq 1}
    \frac{2n+1}{4}
    \left[\frac{\mathcal{N}_{n,2}^{(\mathrm{tr})}\rho^2}{\mathcal{D}_n^{(\mathrm{st})}}+2n\mathcal{F}_n^{(\mathrm{tr})}(\rho, \tau)\right]
    \rho^{n-2}
    P_n(\cos\theta),\\
    \widetilde{\sigma}_{\rho\rho}^{(\mathrm{P},n)}(\rho,\theta,s)
    &\equiv -(2n+1)
    \frac{f_{n,1}(\rho s)f_{n,3}(\mu s)}{\rho^2 s d_n(s)}
    P_n(\cos\theta),
\end{align}
\end{subequations}
where the functions $f_{n,i}(s)$ and $d_n(s)$ are summarized in Tables~\ref{table:functions1} and \ref{table:functions2}.

It should be noted that $\widetilde{\sigma}_{\rho\rho}^{(\mathrm{tr})}$ appears only in the region where the P-wave has arrived but the S-wave has not, and vanishes in the long-time limit $t \to \infty$.
In contrast, $\widetilde{\sigma}_{\rho\rho}^{(\mathrm{P},n)}$ persists even after the arrival of the S-wave and represents the contribution associated with the P-wave.
The corresponding expressions for the displacement components and the other stress components are also listed in Table~\ref{table:functions2}.

\begin{table}[htbp]
    \caption{List of auxiliary functions used in the evaluation of the dynamic solution.}
    \centering
    \begin{tabular}{c}
    \hline
    $\begin{aligned}
    \mathcal{N}_{n,1}^{(\mathrm{tr})}
    &=2(n^2-1)(1-2\nu)\\
    \mathcal{N}_{n,2}^{(\mathrm{tr})}
    &=2(n+1)[2n+1-(2n-1)\nu]\\
    \mathcal{N}_{n,3}^{(\mathrm{tr})}&=2(n^2-1)[1+(2n+1)\nu]\\
    \hline
    \mathcal{A}_n^{(\mathrm{tr})}(\rho) &= 2n(n+2)(1-\nu) + (n^2-1)(1-2\nu)\rho^2\\
    \mathcal{B}_n^{(\mathrm{tr})}
    &=(n^2-1)(2n+3)(1-2\nu)\\
    \mathcal{C}_n^{(\mathrm{tr})}&=3n^3+12n^2+2n-3-(n-1)(4n^2+18n+7)\nu-2(6n+1)\nu^2\\
    \mathcal{F}_n^{(\mathrm{tr})}(\rho, \tau)&= \frac{\mathcal{A}_n^{(\mathrm{tr})}(\rho)+\mathcal{B}_n^{(\mathrm{tr})}\tau^2}{\mathcal{D}_n^{(\mathrm{st})}} - \frac{(n+1)(2n+3)}{2n+5}\frac{\mathcal{C}_n^{(\mathrm{tr})}}{\left(\mathcal{D}_n^{(\mathrm{st})}\right)^2}\\
    \hline
    j_n^{(1)}(s)&= nj_n(s)-sj_{n+1}(s)\\
    f_{n,0}(s)
    &= (n+1)j_n(s) - sj_{n+1}(s)\\
    f_{n,1}(s)
    &= \left[n(n-1)-\frac{1-\nu}{1-2\nu}s^2\right]j_n(s)
    + 2s j_{n+1}(s)\\
    f_{n,2}(s)
    &= (n-1)j_n(s) - sj_{n+1}(s)\\
    f_{n,3}(s)
    &= \left(n^2-1-\frac{s^2}{2}\right)j_n(s) 
    + sj_{n+1}(s)\\
    f_{n,4}(s)
    &= \left(n^2+\frac{\nu}{1-2\nu}s^2\right)j_n(s)
    + sj_{n+1}(s)\\
    f_{n,5}(s)
    &= \left(n-\frac{\nu}{1-2\nu}s^2\right)j_n(s)
    - sj_{n+1}(s)\\
    \hline
    g_{n,1}(s)&= \frac{1}{s}\left[n^2(n-1)-\frac{n-(n-2)\nu}{1-2\nu}s^2\right]j_n(s)
    - \left(n^2+n+2-\frac{1-\nu}{1-2\nu}s^2\right)j_{n+1}(s)\\
    g_{n,2}(s)&=\frac{1}{s}\left[n(n-1)-s^2\right]j_n(s) + 2j_{n+1}(s)\\
    g_{n,3}(s)&= \frac{1}{s}n\left(n^2-1-\frac{s^2}{2}\right)j_n(s)-\left[n(n+1)-\frac{s^2}{2}\right]j_{n+1}(s)\\
    d_n(s)&=g_{n,1}(s)f_{n,3}(\mu s) + \mu f_{n,1}(s)g_{n,3}(\mu s)\\
    &\hspace{1em}
    - n(n+1)\mu g_{n,2}(s)f_{n,2}(\mu s) - n(n+1)f_{n,2}(s)g_{n,2}(\mu s)
    \end{aligned}$\\
    \hline
    \end{tabular}
    \label{table:functions1}
\end{table}
\begin{table}[htbp]
    \caption{List of auxiliary functions used in the evaluation of the dynamic solution (continued).}
    \centering
    \begin{tabular}{c}
    \hline
    $\begin{aligned}
    \widetilde{u}_\rho^{(\mathrm{tr})}(\rho, \theta, \tau)
    &= -\sum_{n\neq 1}\frac{2n+1}{8(n-1)}\left[\frac{N_{n,1}^{(\mathrm{tr})}\rho^2}{D_n^{(\mathrm{st})}}+n \mathcal{F}_n^{(\mathrm{tr})}(\rho, \tau)\right]
    \rho^{n-1} P_n(\cos\theta)\\
    \widetilde{u}_\theta^{(\mathrm{tr})}(\rho,\theta,\tau)
    &= \sum_{n\neq 1}\frac{2n+1}{8(n-1)}\mathcal{F}_n^{(\mathrm{tr})}(\rho, \tau)\rho^{n-1}\sin\theta P_n^\prime(\cos\theta)\\
    \widetilde{\sigma}_{\rho\rho}^{(\mathrm{tr})}(\rho,\theta,\tau)
    &= -\sum_{n\neq 1}
    \frac{2n+1}{4}
    \left[\frac{\mathcal{N}_{n,2}^{(\mathrm{tr})}\rho^2}{\mathcal{D}_n^{(\mathrm{st})}}+n\mathcal{F}_n^{(\mathrm{tr})}(\rho, \tau)\right]\rho^{n-2}
    P_n(\cos\theta)\\
    \widetilde{\sigma}_{\rho\theta}^{(\mathrm{tr})}(\rho,\theta,\tau)
    &= \sum_{n\neq 1}
    \frac{2n+1}{4}
    \mathcal{F}_n^{(\mathrm{tr})}(\rho, \tau)\rho^{n-2}
    \sin\theta P_n^\prime(\cos\theta)\\
    \widetilde{\sigma}_{\theta\theta}^{(\mathrm{tr})}(\rho,\theta,\tau)
    &= -\sum_{n\neq 1}
    \frac{2n+1}{4(n-1)}\rho^{n-2}\nonumber\\
    &\hspace{1em}\times
    \left\{\frac{\mathcal{N}_{n,3}^{(\mathrm{tr})}\rho^2}{\mathcal{D}_n^{(\mathrm{st})}}P_n(\cos\theta)
    - \mathcal{F}_n^{(\mathrm{tr})}(\rho, \tau)\left[n^2P_n(\cos\theta) - \cos\theta P_n^\prime(\cos\theta)\right]\right\}\\
    \widetilde{\sigma}_{\phi\phi}^{(\mathrm{tr})}(\rho,\theta,\tau)
    &= -\sum_{n\neq 1}
    \frac{2n+1}{4(n-1)}\rho^{n-2}\nonumber\\
    &\hspace{1em}\times
    \left\{\frac{\mathcal{N}_{n,3}^{(\mathrm{tr})}\rho^2}{\mathcal{D}_n^{(\mathrm{st})}}P_n(\cos\theta)
    + \mathcal{F}_n^{(\mathrm{tr})}(\rho, \tau)\left[n^2P_n(\cos\theta) - \cos\theta P_n^\prime(\cos\theta)\right]\right\}\\
    \hline
    \widetilde{u}_\rho^{(\mathrm{P}, n)}(\rho, \theta, s)
    &= -\frac{2n+1}{2}\frac{j_n^{(1)}(\rho s)f_{n,3}(\mu s)}{\rho s d_n(s)}P_n(\cos\theta)\\
    \widetilde{u}_\theta^{(\mathrm{P}, n)}(\rho, \theta, s)
    &= \frac{2n+1}{2}\frac{j_n(\rho s)f_{n,3}(\mu s)}{\rho s d_n(s)}\sin\theta P_n^\prime(\cos\theta)\\
    \widetilde{\sigma}_{\rho\rho}^{(\mathrm{P}, n)}(\rho, \theta, s)
    &= -(2n+1)\frac{f_{n,1}(\rho s)f_{n,3}(\mu s)}{\rho^2 s d_n(s)}P_n(\cos\theta)\\
    \widetilde{\sigma}_{\rho\theta}^{(\mathrm{P}, n)}(\rho, \theta, s) 
    &= (2n+1)\frac{f_{n,2}(\rho s)f_{n,3}(\mu s)}{\rho^2 s d_n(s)}\sin\theta P_n^\prime(\cos\theta)\\
    \widetilde{\sigma}_{\theta\theta}^{(\mathrm{P}, n)}(\rho, \theta, s)
    &= (2n+1)\left[\frac{f_{n,4}(\rho s)f_{n,3}(\mu s)}{\rho^2 s d_n(s)}P_n(\cos\theta)
    - \frac{j_n(\rho s)f_{n,3}(\mu s)}{\rho^2 s d_n(s)}\cos\theta P_n^\prime(\cos\theta)\right]\\
    \widetilde{\sigma}_{\phi\phi}^{(\mathrm{P}, n)}(\rho, \theta, s)
    &= -(2n+1)\left[\frac{f_{n,5}(\rho s)f_{n,3}(\mu s)}{\rho^2 s d_n(s)}P_n(\cos\theta)
    - \frac{j_n(\rho s)f_{n,3}(\mu s)}{\rho^2 s d_n(s)}\cos\theta P_n^\prime(\cos\theta)\right]\\
    \hline
    \widetilde{u}_\rho^{(\mathrm{S}, n)}(\rho, \theta, s)
    &= \frac{2n+1}{2}\frac{n(n+1)f_{n,2}(s)j_n(\mu\rho s)}{\rho s d_n(s)}P_n(\cos\theta)\\
    \widetilde{u}_\theta^{(\mathrm{S}, n)}(\rho, \theta, s)
    &= -\frac{2n+1}{2}
    \frac{f_{n,2}(s)f_{n,0}(\mu\rho s)}{\rho s d_n(s)}\sin\theta P_n^\prime(\cos\theta)\\
    \widetilde{\sigma}_{\rho\rho}^{(\mathrm{S}, n)}(\rho, \theta, s)
    &= (2n+1)\frac{n(n+1)f_{n,2}(s)f_{n,2}(\mu\rho s)}{\rho^2 s d_n(s)}P_n(\cos\theta)\\
    \widetilde{\sigma}_{\rho\theta}^{(\mathrm{S}, n)}(\rho, \theta, s)
    &= -(2n+1)
    \frac{f_{n,2}(s)f_{n,3}(\mu\rho s)}{\rho^2 s d_n(s)}
    \sin\theta P_n^\prime(\cos\theta)\\
    \widetilde{\sigma}_{\theta\theta}^{(\mathrm{S}, n)}(\rho, \theta, s)
    &= -(2n+1)\left[\frac{n(n+1)f_{n,2}(s)j_n^{(1)}(\mu\rho s)}{\rho^2 s d_n(s)}
    P_n(\cos\theta)
    - \frac{f_{n,2}(s)f_{n,0}(\mu\rho s)}{\rho^2 s d_n(s)}
    \cos\theta P_n^\prime(\cos\theta)\right]\\
    \widetilde{\sigma}_{\phi\phi}^{(\mathrm{S}, n)}(\rho, \theta, s) 
    &= (2n+1)\left[\frac{n(n+1)f_{n,2}(s)j_n^{(1)}(\mu\rho s)}{\rho^2 s d_n(s)}
    P_n(\cos\theta)
    - \frac{f_{n,2}(s)f_{n,0}(\mu\rho s)}{\rho^2 s d_n(s)}
    \cos\theta P_n^\prime(\cos\theta)\right]
    \end{aligned}$\\
    \hline
    \end{tabular}
    \label{table:functions2}
\end{table}

\subsection{Case for $\tau\ge \mu(1-\rho)$}
In this case, the contribution of the S-wave also appears.
From Eq.~\eqref{eq:asymptotic}, both terms can be evaluated by adding contours
in the complex plane that pass through the region with negative real part
while satisfying $|s|\gg 1$.
Carrying out this procedure, we obtain
\begin{align}
    \widetilde{\sigma}_{\rho\rho}(\rho, \theta, \tau)
    = \widetilde{\sigma}_{\rho\rho}^{(\mathrm{st})}(\rho, \theta)
    + \sum_{n\neq 1} \sum_{m=1}^\infty
    &\left[
    \widetilde{\sigma}_{\rho\rho}^{(\mathrm{P},n)}(\rho,\theta,s_{n,m})
    + \widetilde{\sigma}_{\rho\rho}^{(\mathrm{S},n)}(\rho,\theta,s_{n,m})
    \right]\nonumber\\
    &\times 
    \cos(s_{n,m}\tau),
\end{align}
where $\widetilde{\sigma}_{\rho\rho}^{(\mathrm{P},n)}(\rho,\theta,s)$ is identical to that
in the previous case, while the expression for
$\widetilde{\sigma}_{\rho\rho}^{(\mathrm{S},n)}(\rho,\theta,s)$ is given in
Table~\ref{table:functions2}.
This result indicates that not only the P-wave but also the S-wave contributes
to the stress field in this regime.

\subsection{Results}
\begin{figure}[htbp]
    \centering
    \includegraphics[width=0.75\linewidth]{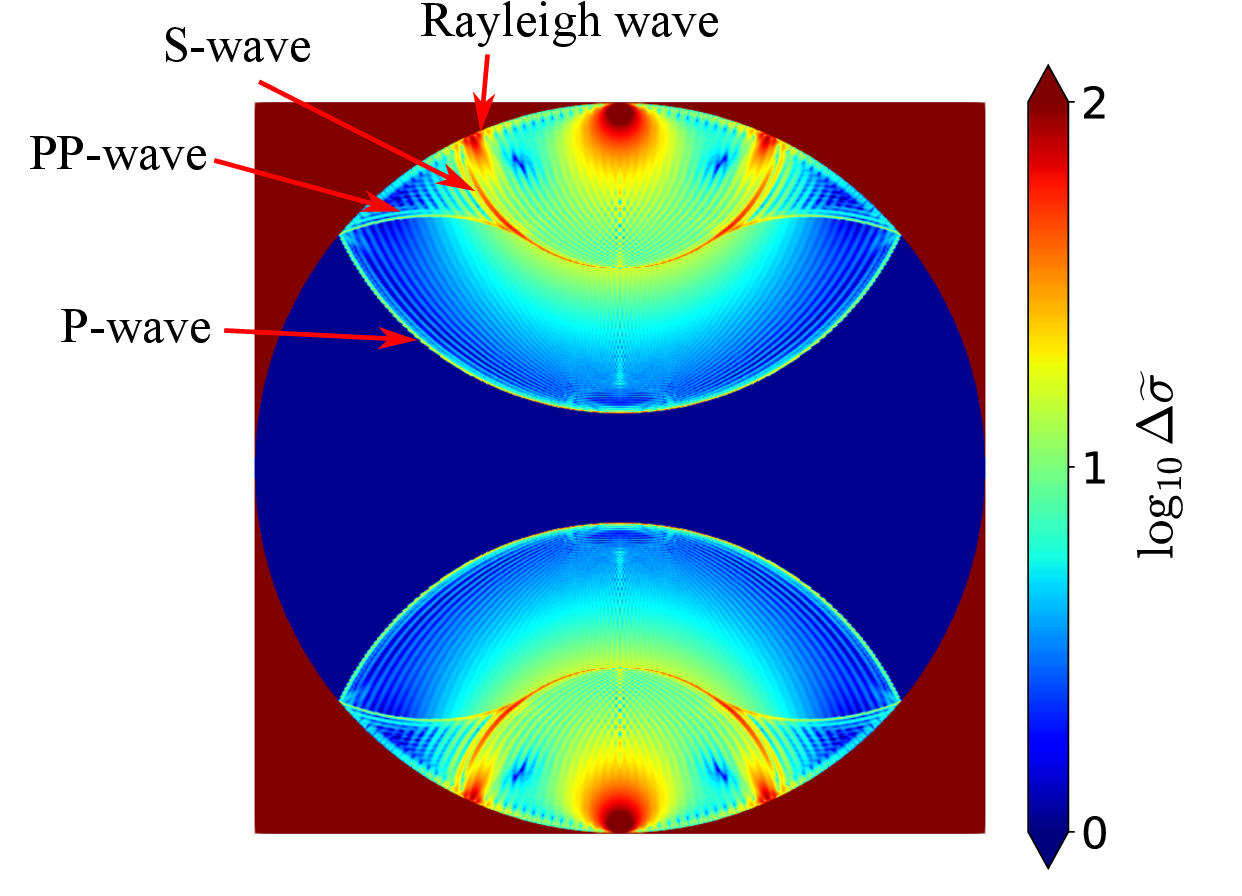}
    \caption{Principal stress difference for $\nu=0.3$ under four concentrated loads: loads of magnitude $\sigma_0$ applied at the north pole ($\theta=0$) and the south pole ($\theta=\pi$).
    The result at $\tau=0.85$ is shown.}
    \label{fig:dynamic}
\end{figure}

Based on the above formulation, the principal stress difference
$\Delta \widetilde{\sigma}$ can be visualized as shown in Fig.~\ref{fig:dynamic}.
As in Ref.~\cite{Sato24_3D}, we consider a loading configuration that satisfies force balance, 
namely, a pair of concentrated loads applied at the north pole ($\theta=0$) and the south pole ($\theta=\pi$). 
Prior to the interaction of waves emitted from the two poles, the solution in each hemisphere 
can be regarded as equivalent to that generated by a single loading point.

In contrast to the static solution presented in Section~\ref{sec:static}, the evaluation of the dynamic solution requires the use of Bessel functions.
Although, in principle, Bessel functions of arbitrary order can be computed, practical implementations in numerical environments such as C or Python are limited to finite orders.
In the present calculation, the summations over $n$ and $m$ are truncated at $200$.
In the static case, the solution is well converged up to the vicinity of the loading point, specifically for $1-\rho \gtrsim 10^{-2}$.
Based on this observation, a similar level of convergence is expected for the dynamic solution.
For a detailed discussion on convergence, see Ref.~\cite{Sato24_3D}.

The results clearly show that both P- and S-waves propagate concentrically from the loading point with dimensionless velocities $1$ and $1/\mu$, respectively, as indicated in Fig.~\ref{fig:dynamic}.
The outermost propagating front corresponds to the P-wave, followed by the S-wave.
In addition, a wave localized near the surface is observed, which corresponds to a Rayleigh wave propagating slightly slower than the S-wave.

A more intricate structure appears after these waves interact with the boundary.
As seen in Fig.~\ref{fig:dynamic}, reflected waves propagate back into the interior, including P-waves reflected at the surface (hereafter referred to as PP-waves).
A notable feature is that these reflected wavefronts do not form simple circular fronts, but instead appear as envelope-like structures.
This behavior can be understood in essentially the same way as in the two-dimensional case~\cite{Sato24_2D}.
As described in Ref.~\cite{Sato24_3D}, a P-wave emitted at $\tau=0$ propagates linearly and reaches a point on the surface at time $\tau'(<\tau)$, namely $(1, \cos^{-1}(1-\tau^{\prime 2}/2))$.
From this point, secondary waves are generated, including reflected P-waves (PP-waves) and mode-converted S-waves.
The observed wavefront is then given by the envelope formed through the superposition of contributions from all emission times $0<\tau'<\tau$.
This explains the characteristic curved structures seen in Fig.~\ref{fig:dynamic}, particularly in the regions labeled as PP-waves.
The loci of the wavefront corresponding to each $\tau'$ are given by
\begin{equation}
    (\rho,\theta)=
    \begin{cases}
        \left(\rho, \cos^{-1}\left(1-\dfrac{\tau^{\prime2}}{2}\right)
        +\cos^{-1}\left[\dfrac{1+\rho^2+(\tau-\tau^\prime)^2}{2\rho}\right]\right)\\
        \hspace{20em} (\text{PP-wave})\\
        \left(\rho, \cos^{-1}\left(1-\dfrac{\tau^{\prime2}}{2}\right)
        +\cos^{-1}\left[\dfrac{\mu^2(1+\rho^2)+(\tau-\tau^\prime)^2}{2\rho}\right]\right)\\
        \hspace{20em} (\text{PS-wave})
    \end{cases}.
\end{equation}
A similar argument applies to waves generated by incident S-waves, leading to additional reflected and mode-converted components.

These results demonstrate that the present formulation is capable of capturing the essential features of wave propagation, including the separation of P- and S-wave contributions, thereby complementing previous transient analyses~\cite{Jingu85_3D, Sato24_3D}.

\section{Conclusion}\label{sec:conclusion}
In this study, we have derived a complete analytical solution for the stress field inside a homogeneous, isotropic solid sphere subjected to a concentrated surface load. 
Starting from the three-dimensional linearized elastodynamic equations, the displacement and stress fields are obtained using scalar and vector potential representations combined with spherical harmonic expansions. 
The static elastic solution is rigorously established as the long-time limit of the dynamical formulation.

The dynamic solution has also been derived, and it has been shown that P- and S-waves propagate through the interior of the sphere with the longitudinal and transverse wave speeds, respectively. 
The analytical expressions further reveal the emergence of Rayleigh waves along the surface as well as reflected waves generated by boundary interactions, whose superposition gives rise to characteristic envelope structures in the stress field.

Closed-form expressions for all components of the stress tensor are derived, enabling direct evaluation of the principal stresses and their differences throughout the interior of the sphere. 
The analysis clarifies the contribution of each spherical harmonic mode and explicitly demonstrates that the $n=1$ mode represents rigid-body translation and does not generate internal elastic stress. 
This resolves an aspect that is often treated implicitly in related analytical treatments.

The principal stress difference, which is of central importance in three-dimensional photoelasticity, was evaluated analytically and its spatial structure was examined in detail. 
The solution provides a rigorous theoretical reference for interpreting photoelastic fringe patterns and for validating numerical or experimental reconstructions of internal stress fields.

By exploiting rotational symmetry, the solution obtained for polar loading was generalized to arbitrary loading positions, and complex loading configurations were treated systematically through superposition. 
The present formulation therefore offers a unified analytical framework for concentrated surface loading in bounded spherical elastic bodies.
Compared with existing formulations, the present solution provides explicit expressions for all coefficients, which facilitates direct numerical evaluation and clear physical interpretation, particularly in the dynamic regime.

Although the analysis is restricted to linear elasticity and idealized concentrated tractions, the framework developed here can be extended to distributed loads or more complex boundary conditions. 
We expect that the analytical solution presented in this work will serve both as a benchmark for computational methods and as a theoretical foundation for three-dimensional stress visualization and experimental mechanics.

\section*{Acknowledgement}
The author gratefully acknowledges Yosuke Sato for providing a prototype code for the numerical computation of the dynamical solution.
The authors also thank the anonymous reviewers for their valuable comments and suggestions, which have helped improve the manuscript.

\section*{Funding}
This work is partially supported by the Grant-in-Aid of MEXT for Scientific Research (Grant No.~JP24K06974, No.~JP24K07193, No.~JP24KJ0110, and No.~JP25K01063).

\appendix
\section{Comparison with Finite Element Method}\label{sec:FEM}
In this appendix, we validate the theoretical results presented in this paper by comparing them with numerical results obtained using the finite element method (FEM)~\cite{Hughes12}. 
Strictly speaking, the present analytical solution can serve as a benchmark for assessing the accuracy of FEM. 
Nevertheless, it is still meaningful to first confirm the agreement between the two approaches under a relatively simple loading condition.

Here, we consider the case in which concentrated loads are applied at the north pole ($\theta = 0$) and the south pole ($\theta = \pi$) of an elastic sphere. 
The FEM analysis is carried out using MATLAB's built-in finite element solver. 
The computational mesh is automatically generated, and no manual refinement is introduced. 
The boundary conditions are identical to those used in the theoretical analysis.

\begin{figure}[htbp]
    \centering
    \includegraphics[width=0.75\linewidth]{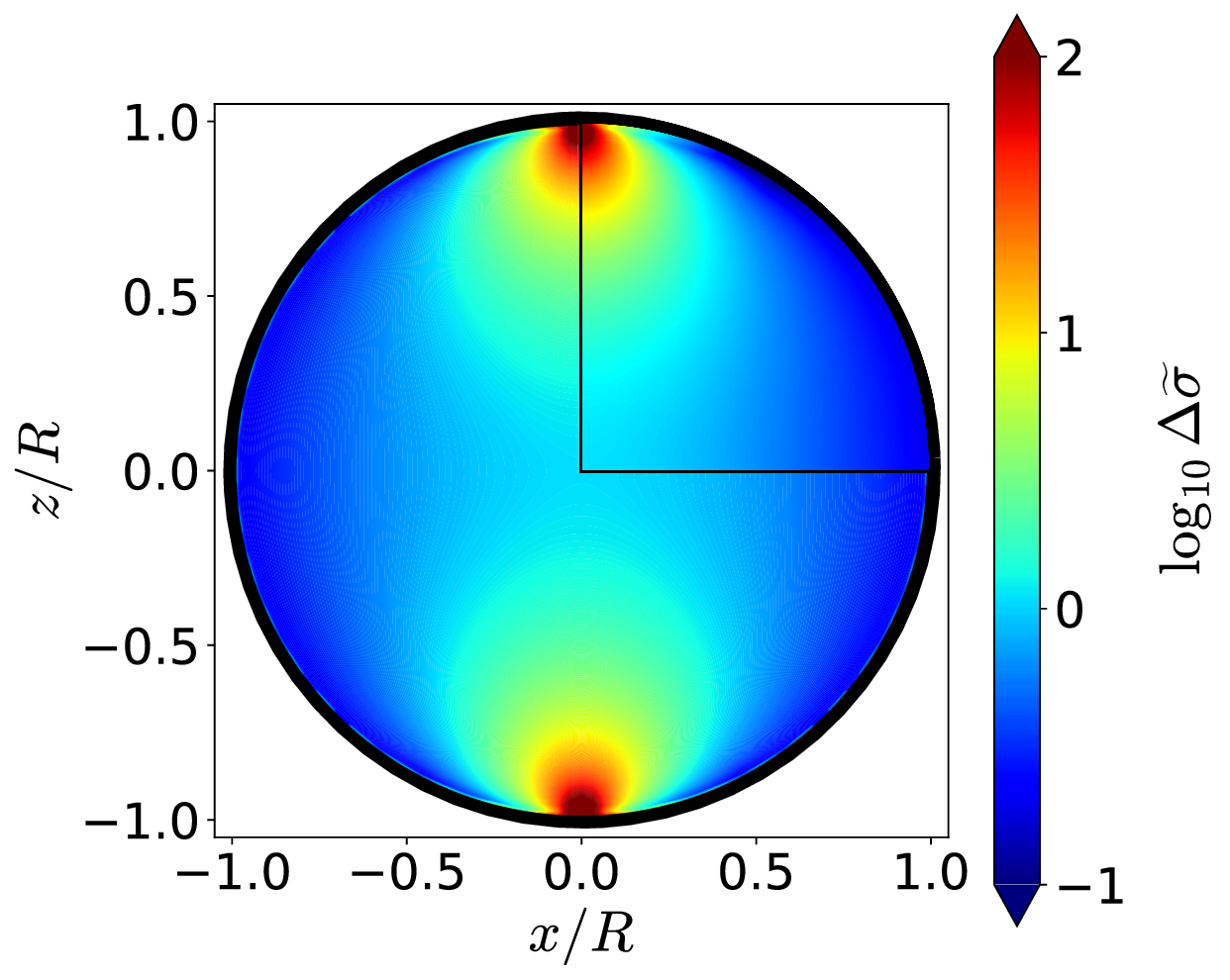}
    \caption{Principal stress difference for $\nu=0.3$ under four concentrated loads: loads of magnitude $\sigma_0$ applied at the north pole ($\theta=0$) and the south pole ($\theta=\pi$).
    The upper-right quadrant shows the FEM result.}
    \label{fig:FEM}
\end{figure}

Figure~\ref{fig:FEM} presents a comparison between the analytical and FEM results. 
Overall, the two are in good agreement, supporting the validity of the theoretical solution. 
However, slight distortions in the contours of the principal stress difference are observed in the FEM results near the surface. 
These artifacts are likely attributable to insufficient mesh resolution and the inherent difficulty of accurately representing concentrated loads within the finite element framework. 
A finer mesh or more sophisticated treatment of the applied loads would be expected to mitigate these discrepancies.


\end{document}